\documentclass{pas}

\usepackage{graphicx}
\usepackage{xcolor}
\usepackage{url}

\usepackage[normalem]{ulem}

\usepackage[nolist]{acronym}
\begin{acronym}[]
    \acrodef{UEMR}{unintended electromagnetic radiation}
    \acrodef{SKA}{Square Kilometre Array}
    \acrodef{ATNF}{Australia Telescope National Facility}
    \acrodef{ATCA}{Australia Telescope Compact Array}
    \acrodef{DTC}{Direct-to-Cell}
    \acrodef{DTD}{Direct-to-device}
    \acrodef{BA}{boresight avoidance}
    \acrodef{ODS}{Observational Data Sharing}
    \acrodef{DTH}{Direct-to-Handset}
    \acrodef{DTM}{Direct-to-mobile}
    \acrodef{SNIFFLES}{Satellite measuremeNt of Intended emission, unwanted emission, and radio radiation to develop, Follow up, and veriFy mitigation measures, regulatory compliance, and Lawful usE of the radio Spectrum}
    \acrodef{NGSO}{non-geostationary satellite orbit}
    \acrodef{LEO}{low-Earth orbit}
    \acrodef{MEO}{medium-Earth orbit}
    \acrodef{EOP}{Earth Orientation Parameter}
    \acrodef{GNSS}{Global Navigation Satellite Systems}
    \acrodef{EMI}{electromagnetic interference}
    \acrodef{EMC}{electromagnetic compatibility}
    \acrodef{OOBE}{out-of-band emission}
    \acrodef{NRAO}{National Radio Astronomy Observatory}
    \acrodef{CABB}{Compact Array Broadband Backend}
    \acrodef{BIGCAT}{Broadband Integrated GPU Correlator for ATCA}
    \acrodef{LOFAR}{LOw-Frequency ARray}
    \acrodef{RAS}{radio astronomy service}
    \acrodef{ITU-R}{International Telecommunication Union Radiocommunication Sector}
    \acrodef{RR}{Radio Regulations}
    \acrodef{ASKAP}{Australian SKA Pathfinder}
    \acrodef{VLBI}{very long baseline interferometry}
    \acrodef{VGOS}{VLBI Global Observing System}
    \acrodef{FSS}{Fixed Satellite Service}
    \acrodef{RFI}{radio-frequency interference}
    \acrodef{e.i.r.p}{effective isotropic radiated power}
    \acrodef{EPFD}{equivalent power flux density}
    \acrodef{SKAO}{Square Kilometre Array Observatory}
    \acrodef{MWA}{Murchison Widefield Array}
    \acrodef{ARQZWA}{Australian Radio Quiet Zone Western Australia}
    \acrodef{GFS}{NOAA Global Forecast System}
    \acrodef{RRL}{ACMA Register of Radiocommunications Licences}
    \acrodef{ACMA}{Australian Communications and Media Authority}
    \acrodef{ADS-B}{Automatic Dependent Surveillance--Broadcast}
    \acrodef{AIS}{Automatic Identification System}
    \acrodef{AMSA}{Australian Maritime Safety Authority}
    \acrodef{AUC}{area under the receiver operating characteristic curve}
    \acrodef{BSS}{Brier skill score}
    \acrodef{DEM}{digital elevation model}
    \acrodef{DTV}{digital television}
    \acrodef{DVB-T}{Digital Video Broadcasting--Terrestrial}
    \acrodef{ECMWF}{European Centre for Medium-Range Weather Forecasts}
    \acrodef{EIRP}{effective isotropic radiated power}
    \acrodef{ENSO}{El Ni\~no--Southern Oscillation}
    \acrodef{FM}{frequency modulation}
    \acrodef{GPS}{Global Positioning System}
    \acrodef{IMT}{International Mobile Telecommunications}
    \acrodef{LTE}{Long-Term Evolution}
    \acrodef{MAD}{median absolute deviation}
    \acrodef{MMSI}{Maritime Mobile Service Identity}
    \acrodef{NWP}{numerical weather prediction}
    \acrodef{SRTM}{Shuttle Radar Topography Mission}
    \acrodef{UHF}{ultra-high frequency}
    \acrodef{VHF}{very high frequency}
    \acrodef{RALI}{Radiocommunications Assignment and Licensing Instruction}
    \acrodef{GBM}{gradient-boosted model}
\end{acronym}

\usepackage{remreset}
\makeatletter
\@ifundefined{c@figure}{}{}
\makeatother

\begin{document}

\lefttitle{Publications of the Astronomical Society of Australia}
\righttitle{Tropospheric Ducting Prediction at Inyarrimanha Ilgari Bundara}

\jnlPage{1}{1}
\jnlDoiYr{2026}
\doival{10.1017/pasa.xxxx.xx}

\articletitt{Research Paper}

\title{Tropospheric Ducting Prediction based on GFS Model Data at Inyarrimanha Ilgari Bundara, the CSIRO Murchison Radio-astronomy Observatory}

\author{\gn{Balthasar} \sn{Indermuehle}$^{1}$ and
        \gn{Hajime} \sn{Suzuki}$^{2}$
}

\affil{$^1$CSIRO Space \& Astronomy, PO Box 76, Epping NSW 1710, Australia\\
       $^2$CSIRO Technology, PO Box 76, Epping NSW 1710, Australia}

\corresp{B. Indermuehle, Email: balt.indermuehle@csiro.au}

\citeauth{Indermuehle B and Suzuki H, Tropospheric ducting prediction based on GFS model data at Inyarrimanha Ilgari Bundara, the CSIRO Murchison Radio-astronomy Observatory. {\it Publications of the Astronomical Society of Australia}
{\bf 00}, 1--12. https://doi.org/10.1017/pasa.xxxx.xx}

\history{(Received xx xx xxxx; revised xx xx xxxx; accepted xx xx xxxx)}

\begin{abstract}
Tropospheric ducting transports terrestrial \ac{RFI} over hundreds of kilometres into the \ac{ARQZWA} at Inyarrimanha Ilgari Bundara, the CSIRO Murchison Radio-astronomy Observatory. We present a ducting prediction method based on vertical refractivity profiles derived from \ac{GFS} 0.25$^{\circ}$ data, validated against seven years of continuous spectrum monitoring at 25 channels between 88 MHz and 2.68 GHz. Candidate interference sources are attributed from the \ac{RRL}. Per-transmitter basic transmission loss to the site is computed with Recommendation ITU-R P.452-16 above 100 MHz and Recommendation ITU-R P.1812-6 below it, over Copernicus GLO-90 terrain, driven at each timestep by the GFS-derived refractivity lapse along the path. Skill is scored as a stratified \ac{AUC}, computed within month and three-hour strata so that the seasonal and diurnal cycle shared with the predictor cannot inflate it. Across the 25 channels the median stratified \ac{AUC} is 0.724 and 24 of 25 confidence intervals exclude chance. A time-percentage calibration derived from four pilot months transfers without retuning to the full record, and skill is flat through three days of forecast lead. A pooled \ac{GBM} on the same physical features, augmented by a duct-corridor connectivity metric, reaches a mean test \ac{AUC} of 0.806 over all 25 channels and 0.799 over the 16 on which the analytical predictor is itself skilled. In-situ \ac{LTE} cell decoding confirms attribution: 86--97\% of over-the-horizon cell detections fall within flagged event hours, the range spanning the four monitored \ac{LTE} channels, and the decoded network identities match the licensing records. Ship \ac{AIS} receptions at 162 MHz, each a self-located 12 W transmitter at a known sea position, provide \ac{VHF} ground truth: reception paths show a 67\% higher land-segment duct fraction than matched controls. Two mobile channels at 2.6 GHz were recovered from self-generated site interference by discriminating on bandwidth rather than amplitude, a method transferable to any polluted observatory monitoring band. The \ac{FM} channels retain only weak skill, far below the ducting channels, and we attribute their events to near-threshold local sources after excluding ducting, sporadic-E and aircraft scatter; the 460 MHz land mobile channel is the one well-powered null in the set. The method enables adaptive scheduling of observations away from frequencies affected by forecast ducting, and drives a live forecast service at the observatory.
\end{abstract}


\maketitle
\acresetall

\section{Introduction}
Inyarrimanha Ilgari Bundara, the CSIRO Murchison Radio-astronomy Observatory, hosts the \ac{ASKAP}, the \ac{MWA}, and the low-frequency component of the \ac{SKAO}, currently under construction. The site was chosen for its radio quietness \citep{bowman2007} and is protected by the \ac{ARQZWA} \citep{wilson2016, indermuehle2016, wilson2011}, implemented through \ac{RALI} MS32 \citep{acma2014}. Population density in the surrounding Murchison shire is among the lowest in Australia, and the nearest sizeable population centres, Geraldton and Perth, lie approximately 300 km and 600 km away \citep{tingay2020, sokolowski2016}, far beyond the radio horizon. The \ac{ARQZWA} restricts licensed transmitters within a 70 km inner zone, applies protection criteria in an outer zone to 150 km, and requires coordination in frequency-dependent zones extending to 260 km from the site \citep{acma2014, wilson2011, itu_ra2259}. The protection thresholds derive from the radio astronomy interference criteria of Recommendation ITU-R RA.769 \citep{itu_ra769}.

Tropospheric ducting episodically defeats this isolation. The radio refractivity $N$ of air is set by pressure, temperature, and water vapour pressure \citep{smith1953, itu_p453_14}; in a standard atmosphere it decreases by about 40 N-units per km of height. Temperature inversions and sharp humidity lapses can steepen this gradient beyond $-157$ N-units per km into the \emph{super-refractive} domain. The modified refractivity $M = N + 0.157z$ (with $z$ in metres) then decreases with height, and the layer traps radio waves, guiding them along the Earth's curvature \citep{kerr1951, bean1966, itu_p310_11}. Signals in the \ac{VHF} and \ac{UHF} bands coupled into such a super-refractive duct propagate for hundreds of kilometres with losses far below free-space-plus-diffraction expectations \citep{turton1988, hitney1985}. The responsible meteorology, subsidence inversions, nocturnal radiation inversions, and advection of maritime air over land, is common in subtropical latitudes and peaks in the warm months \citep{turton1988, vonengeln2004}.

Ducted terrestrial transmissions are a documented interference source at the observatory. \ac{MWA} observations show \ac{DTV} signals present 3\% of the time within ten observation days, attributed to occasional ionospheric or atmospheric propagation \citep{offringa2015}. All-sky monitoring with BIGHORNS identified 98 tropospheric ducting events, mostly \ac{DTV} from Perth transmitters some 600 km distant, in nearly two years of data, affecting almost 20\% of analysed nights with a summer maximum \citep{sokolowski2015, sokolowski2016}. \ac{ASKAP} commissioning observations with BETA detected transmitters from below the radio horizon via ducting across the 0.7--1.8 GHz band \citep{indermuehle2016}. Imaging with an SKA-Low prototype station resolved individual \ac{FM} transmitters in Geraldton \citep{tingay2020}, and the transient monitor on SKA-Low prototype stations routinely classifies over-the-horizon \ac{FM} signals \citep{sokolowski2021}. These events raise the noise floor, can saturate sensitive receiver chains, and contaminate science data across the full frequency range of the site's instruments. The ability to predict ducting allows adaptive scheduling of observations at frequencies least affected by the anticipated ducting-facilitated interference.

Ducting conditions are predictable from the same thermodynamic profiles that \ac{NWP} models forecast. Global duct climatologies have been derived from \ac{ECMWF} analysis fields \citep{vonengeln2004, lopez2009} and from \ac{GPS} radio occultation \citep{ao2007, xie2010, feng2020}. Mesoscale models reproduce observed coastal trapping layers with useful skill \citep{burk1997, haack2001, atkinson2001, atkinson2006}. Reanalysis-based regional climatologies \citep{sirkova2015, cheng2021, zhou2022} and machine-learning duct predictors \citep{huang2022, chai2023} extend this work, and weather radar anomalous propagation provides an independent observational record of ducting occurrence \citep{peter2014, norin2023}. None of this work, however, connects \ac{NWP} duct diagnostics to measured interference at a radio-quiet observatory, and no published system predicts ducting-borne interference for observatory operations.

This paper closes that gap. We derive vertical refractivity profiles from the 0.25$^{\circ}$ \ac{GFS} \citep{gfs_ncep, gfs_aws}, compute duct diagnostics along transmitter-to-site paths, and model per-transmitter basic transmission loss over Copernicus GLO-90 terrain \citep{copernicus_dem} with Recommendation ITU-R P.452-16 \citep{itu_p452_16}, which includes ducting and layer reflection among its clear-air mechanisms, as implemented in the pycraf package \citep{pycraf_ascl, winkel2018}; channels below 100 MHz use Recommendation ITU-R P.1812-6 \citep{itu_p1812_6, py1812}. Candidate transmitters are drawn from the \ac{ACMA} database, the \ac{RRL} \citep{acma_rrl}. A note on terminology: the quantity we compute is the \emph{basic transmission loss} of Recommendation ITU-R P.341 \citep{itu_p341}, the loss between isotropic antennas, because that is what the ITU-R propagation recommendations return; it is the same quantity usually called \emph{path loss} in the radio astronomy literature, and we use the ITU-R term throughout for consistency with the models it comes from. Two distinct prediction layers are reported. The first is deterministic: the ITU-R recommendations above, evaluated on the GFS-derived refractivity along each path, with no parameter fitted to the monitoring record. Sections~\ref{sec:validation} and \ref{sec:discussion} score that layer on its own, and it carries the physical claims of the paper. The second is statistical, and measures how much skill the same physical inputs can still support: a single gradient-boosted decision-tree ensemble, pooled across channels rather than fitted per channel, trained on a strictly earlier period than it is tested on. Its inputs fall into three groups:
\begin{itemize}
    \item The analytical ITU-R prediction itself.
    \item The raw duct diagnostics along the path.
    \item A diurnal--annual climatology encoded as sine and cosine harmonics of hour-of-day and day-of-year.
\end{itemize}
The climatology harmonics are genuine input features, and the same ensemble restricted to them alone is the reference against which the physical features are scored in Section~\ref{sec:ml}. We validate the predictions against continuous spectrum monitoring at 25 channels between 88 MHz and 2.68 GHz. Section~\ref{sec:observations} describes the monitoring system and the interference event definition. Section~\ref{sec:gfs} presents the GFS-derived refractivity diagnostics. Section~\ref{sec:attribution} attributes candidate interference sources. Section~\ref{sec:pathloss} details the basic transmission loss modelling. Section~\ref{sec:validation} validates predictions against observed events. Sections~\ref{sec:discussion} and \ref{sec:conclusions} discuss operational implications and conclude.

\section{Ducted signal observations at the observatory}
\label{sec:observations}

\subsection{Monitoring system}
The observatory's \ac{RFI} monitoring system receives through a Rohde \& Schwarz HE600 active antenna mounted on the site \ac{RFI} tower (latitude $-26.6902^{\circ}$, longitude $+116.6238^{\circ}$) at 21 m above ground. An RF-over-fibre link carries the signal to the receivers in the observatory control building. The spectrum monitor sweeps 20--3000 MHz every 62 s and records the peak received power in each monitored channel; we use 30-minute maxima throughout. Co-sited instruments on the same antenna feed include a \ac{LTE} cell scanner, a marine \ac{AIS} decoder, and an aviation \ac{ADS-B} receiver; these provide independent attribution evidence used in Sections~\ref{sec:validation} and \ref{sec:discussion}.

Twenty-five channels are monitored, spanning 88 MHz to 2.68 GHz, all normally clean apart from the two 2.6 GHz channels of Section~\ref{sec:imt}: five in the \ac{FM} broadcast band (88, 88.3, 89.3, 92.3, 97 MHz) and two in the marine \ac{AIS} band (162, 162.02 MHz), four \ac{DVB-T} television channels (202, 536, 564, 631 MHz), one land mobile channel (460 MHz), eight mobile-service downlink channels (763, 778, 872, 882, 947, 955, 1842 MHz and 2650 MHz), one fixed link (1500 MHz), two private \ac{LTE} channels (1865, 2125 MHz), one fixed wireless channel (2351 MHz), and a second \ac{IMT} channel at 2680 MHz. The record runs from 2019 April to 2026 July. \ac{GFS} changed model version on 2021 March 22, from v15 to v16; the v16 era is the analysis span, 44\,255 hours per channel, and the earlier v15 era, 5\,274 hours per channel, is held out entirely as an out-of-era validation block (Section~\ref{sec:validation}). Under normal propagation every channel sits at the receiver noise floor: all licensed co-channel transmitters lie beyond the radio horizon.

\subsection{Event definition and seasonality}
A channel's excess is its 30-minute maximum minus a centred rolling 7-day median baseline. The channel is flagged as elevated when the excess exceeds $\max(0.6\,\mathrm{dB},\, 6 \times 1.4826 \times \mathrm{MAD})$, where \ac{MAD} is a centred rolling 7-day median of the absolute excess and the factor 1.4826 rescales it to a Gaussian-equivalent standard deviation. The binding term differs by observable. On the 23 swept-power channels the dispersion term always binds: the floor fires on none of the triggers, and removing it would add 0.10\% of events. On the two \ac{IMT} pedestal channels of Section~\ref{sec:imt} the pedestal statistic is quantised, its rolling \ac{MAD} is identically zero in most hours, and the 0.6 dB floor is therefore the operative criterion. Events on those two channels should be read as fixed-threshold excursions, not as six-sigma ones.

\begin{figure}[t]
\includegraphics[width=\columnwidth]{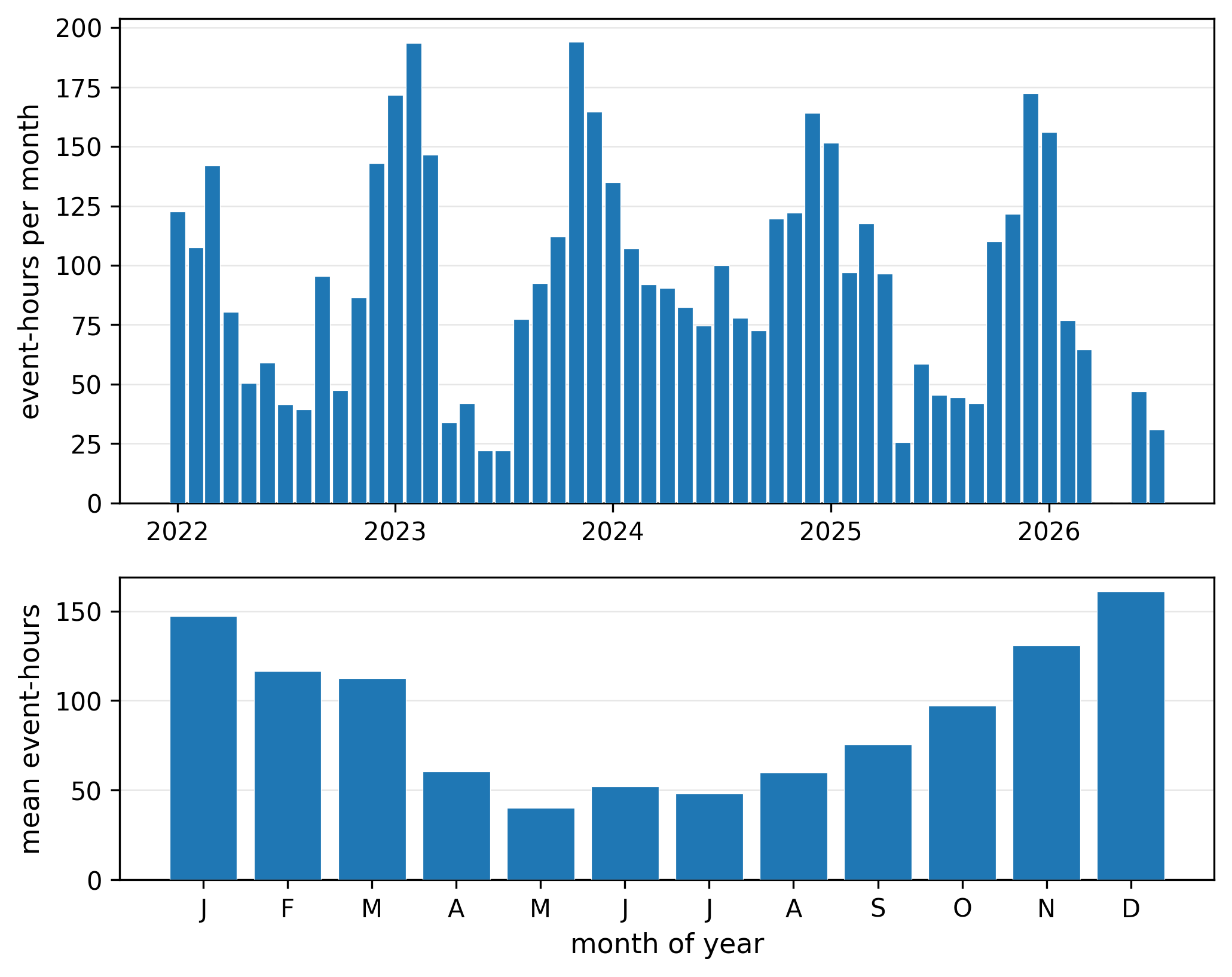}
\caption{Interference event-hours at the observatory under the multi-channel event definition. Top: monthly totals over 2022 January to 2026 July, the span of the multi-channel event catalogue. Bottom: month-of-year means. The summer (November--March) maximum and May--July minimum track the seasonal occurrence of duct-favourable meteorology.}
\label{fig:climatology}
\end{figure}

A candidate ducting period is an hour with at least three channels simultaneously elevated, which suppresses single-channel artefacts such as local transmissions and receiver anomalies. The events are strongly seasonal (Figure~\ref{fig:climatology}). The summer months (November to March) dominate; May to July fall to about a third of the summer rate. This mirrors the summer maximum reported from BIGHORNS all-sky data \citep{sokolowski2016} and the warm-season peak of duct-favourable meteorology \citep{vonengeln2004}.

For calibration we use a pilot period covering the ducting-rich summer quarter 2023 January 1 to March 31 and a quiet control month, 2023 June.

\subsection{Recovering two channels at 2.6 GHz by bandwidth discrimination}
\label{sec:imt}
The 2650 MHz signal band shares spectrum with self-generated \ac{ASKAP} phased-array-feed leakage, a comb of one to two narrow lines. Band power cannot reject a comb, so the band-power detection scored an \ac{AUC} of 0.395 against the duct predictor, below chance, and its event rate tracked the observatory's own \ac{RFI} environment rather than the weather.

The discriminant is bandwidth, not amplitude. A licensed \ac{IMT} downlink is a broadband pedestal filling an entire allocation block; the leakage is narrowband. We therefore take the median of each 1 MHz sub-block minus an in-scan reference and require the \emph{weakest} sub-block to clear threshold, which a comb cannot achieve however strong its individual lines. The blocks are 2632--2648, 2652--2668 and 2672--2688 MHz, with a reference at 2585--2590 MHz. The first two share a tower, established empirically because the licence register carries all of this spectrum as a single nominal 2650.0 MHz entry and the band is spectrum-licensed and therefore invisible in the register; conditioning on hours when all three blocks are elevated, which holds the duct fixed, gives a correlation of 0.80 between the first two blocks against 0.15 between the first and third. The 2650 MHz detection is accordingly the per-scan maximum of the first two blocks, and 2680 MHz is the third. While the discriminant is one of bandwidth alone, it does not assert that every narrowband emission in this band is our own leakage: licensed narrowband services and external narrowband interference would be rejected identically. What it guarantees is the converse, a signal that fills an entire allocation block. This cannot be produced by a comb of narrow lines.

\begin{figure*}
\includegraphics[width=\textwidth]{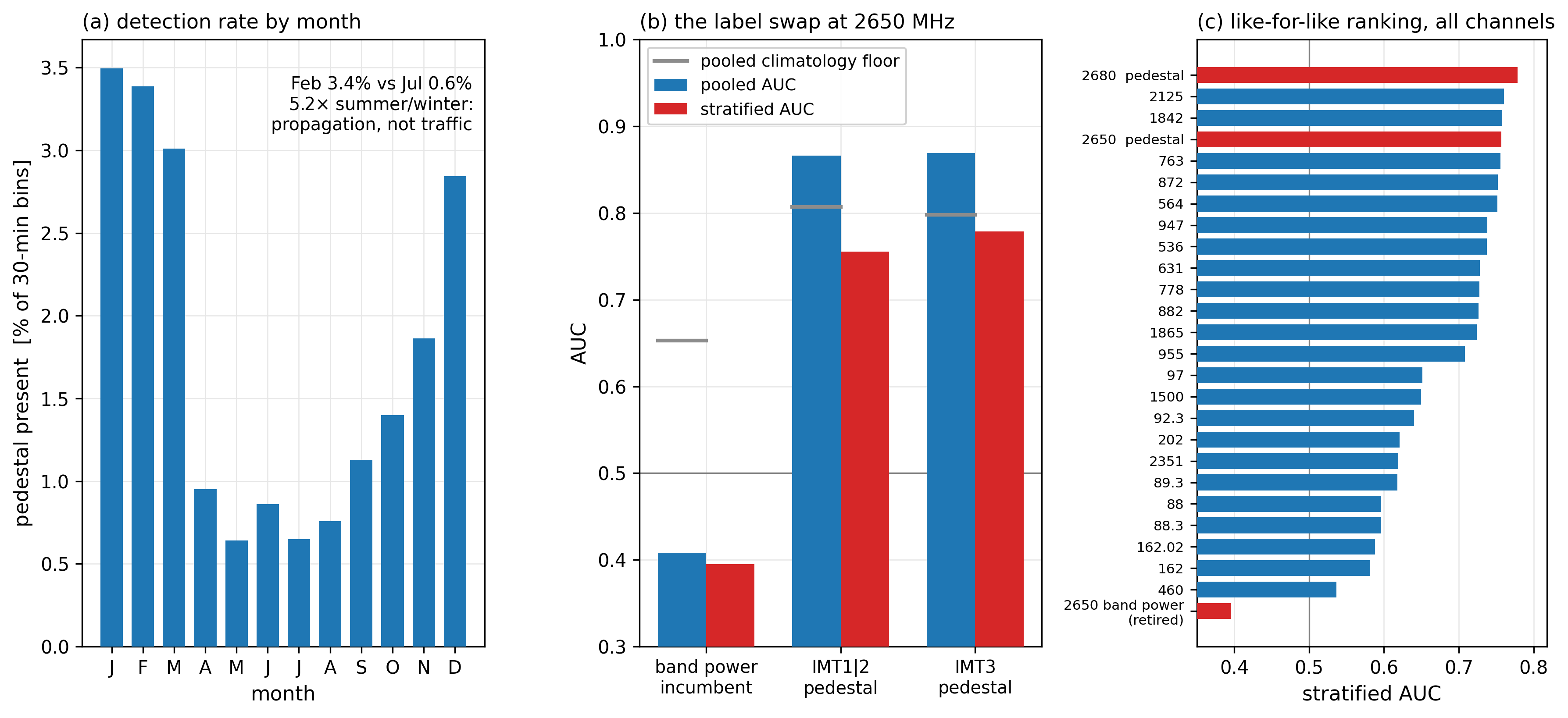}
\caption{Recovering the 2.6 GHz channels. (a) Pedestal detection rate by month, a factor 5.2 between February and July. (b) The label swap at 2650 MHz, with the pooled climatology floor drawn: the pooled bar on the band-power label sits barely above its own climatology floor, so most of it is the shared seasonal cycle rather than event-level skill. (c) Like-for-like ranking of all 25 channels, with the retired band-power label appended at the bottom.}
\label{fig:imtrescue}
\end{figure*}

Predictors were held fixed throughout, so the comparison isolates detection quality: the stratified \ac{AUC} rises from 0.395 on band power to 0.757 on the pedestal detection at 2650 MHz, moving the channel from last to fourth of 25 (Figure~\ref{fig:imtrescue}). The recovered detections carry a strong seasonal cycle. At the detector's own threshold the pedestal is present in 3.39\% of February scan bins against 0.65\% in July, a ratio of 5.2; restricting to the single narrowest allocation block, which is the stricter test, the same months read 1.42\% and 0.22\%, a ratio of 6.4. Licensed traffic would not sharpen with the detection threshold, so the seasonality is propagation rather than usage. The method is transferable. Any observatory monitoring a band that it also pollutes can separate licensed occupancy from its own leakage on bandwidth, without needing to model or subtract the interferer.

\section{GFS model data and refractivity diagnostics}
\label{sec:gfs}

Atmospheric state comes from the archival NOAA \ac{GFS} at 0.25$^{\circ}$ resolution, retrieved as GRIB2 from the NOAA Open Data Dissemination archive on AWS \citep{gfs_ncep, gfs_aws}. Forecast hours f000 to f005 of each 6-hourly cycle are concatenated into continuous hourly coverage. Each file contains temperature, relative humidity, and geopotential height on 21 isobaric levels, plus the 2-m and 0.995-sigma near-surface fields, over an Australia-west region spanning latitudes $-36^{\circ}$ to $-20^{\circ}$ and longitudes $109^{\circ}$ to $125^{\circ}$ (a 65$\times$65 grid).

Per grid column we compute the radio refractivity
\begin{equation}
N = 77.6\,\frac{p}{T} + 3.73\times10^{5}\,\frac{e}{T^{2}},
\label{eq:smithweintraub}
\end{equation}
with pressure $p$ and water vapour pressure $e$ in hPa and temperature $T$ in K \citep{smith1953, itu_p453_14}. Vapour pressure follows from relative humidity as $e = (\mathrm{RH}/100)\,e_{s}(T)$, with the saturation vapour pressure $e_{s}$ of \citet{bolton1980}. Profiles are assembled from the surface up: the 2-m fields anchor the base, the 0.995-sigma level sits immediately above, and isobaric levels below the local surface are masked. This is important at the site, which lies at $\sim$370 m elevation with a typical surface pressure of $\sim$975 hPa: the 1000 hPa level is below ground.

\begin{figure*}
\includegraphics[width=\textwidth]{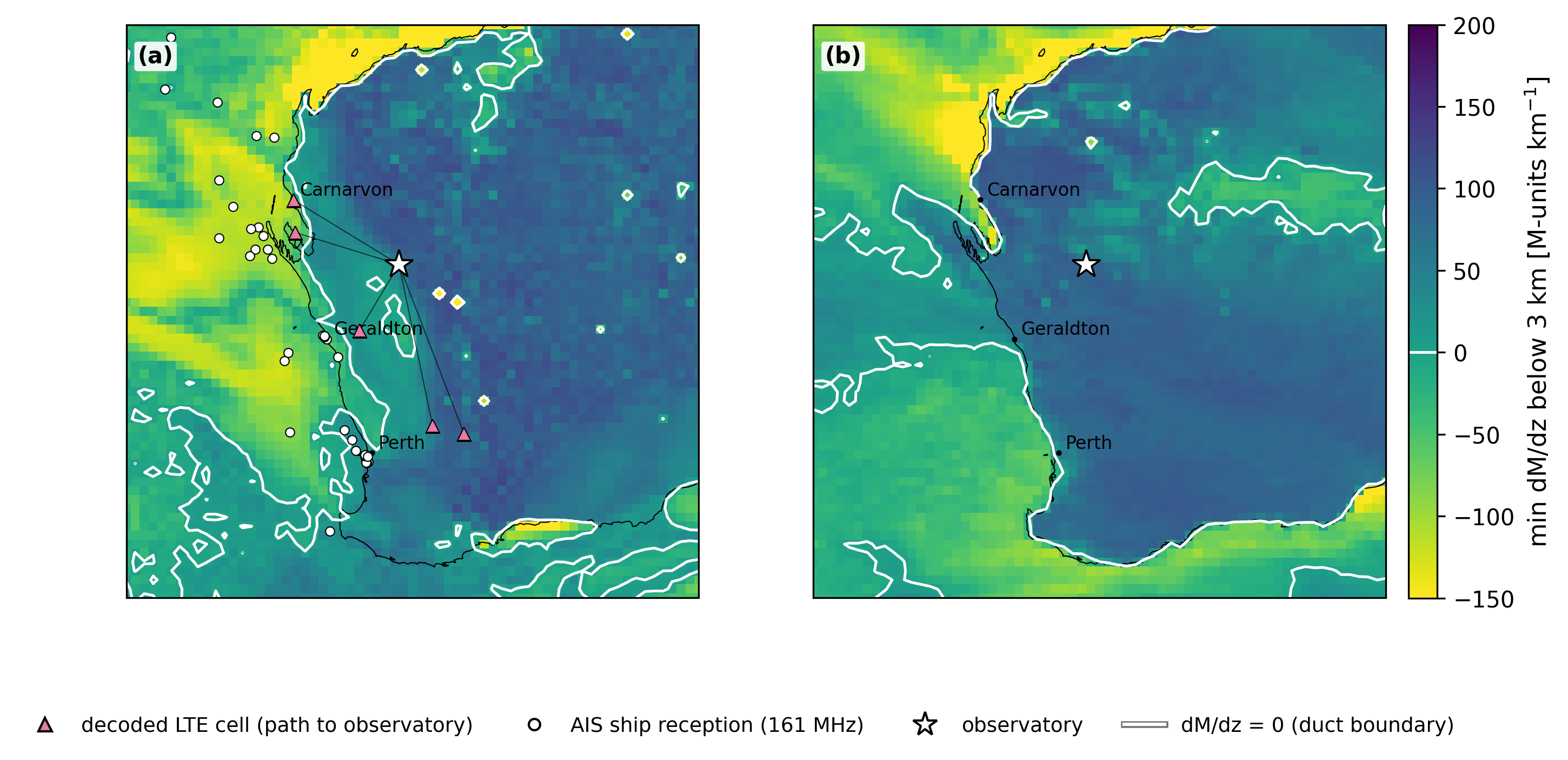}
\caption{Minimum vertical gradient of modified refractivity below 3 km from the GFS-derived profiles. (a) The 2023 February 5 21:00 UT event, during which ten monitored channels were simultaneously elevated. Bright colours indicate trapping; the white contour marks $\mathrm{d}M/\mathrm{d}z = 0$, the duct boundary. Pink triangles are over-the-horizon \ac{LTE} cells decoded at the site during the event, with their great-circle paths to the observatory (star); white circles are \ac{AIS} ship receptions in the surrounding window. (b) A quiet winter control, 2023 June 15 09:00 UT: ducting is confined offshore and the coastal bridge is absent.}
\label{fig:ductcase}
\end{figure*}

The modified refractivity $M = N + 0.157z$ ($z$ in m) then yields the duct diagnostics: the minimum vertical gradient $\mathrm{d}M/\mathrm{d}z$ in the lowest 3 km, the corresponding minimum $\mathrm{d}N/\mathrm{d}z$, the lowest-layer $N$ gradient, the duct base and top heights where $\mathrm{d}M/\mathrm{d}z < 0$, the integrated $M$ deficit across the trapping layer (duct depth), and a binary ducting flag. These are produced for every grid column and hour, giving duct-diagnostic maps of the region as well as profiles along transmitter-to-site paths. Figure~\ref{fig:ductcase} contrasts the region during a strong multi-channel event with a quiet winter control.

\section{Interference source attribution}
\label{sec:attribution}

Candidate interference sources for each monitored channel are drawn from year-matched snapshots of the \ac{RRL}. A licensed emission matches a monitored channel when its emission-designator bandwidth overlaps the monitor channel within a 0.5 MHz margin. Matches are deduplicated by site, so multiple licences at one location count once per channel.

\begin{figure*}
\includegraphics[width=\textwidth]{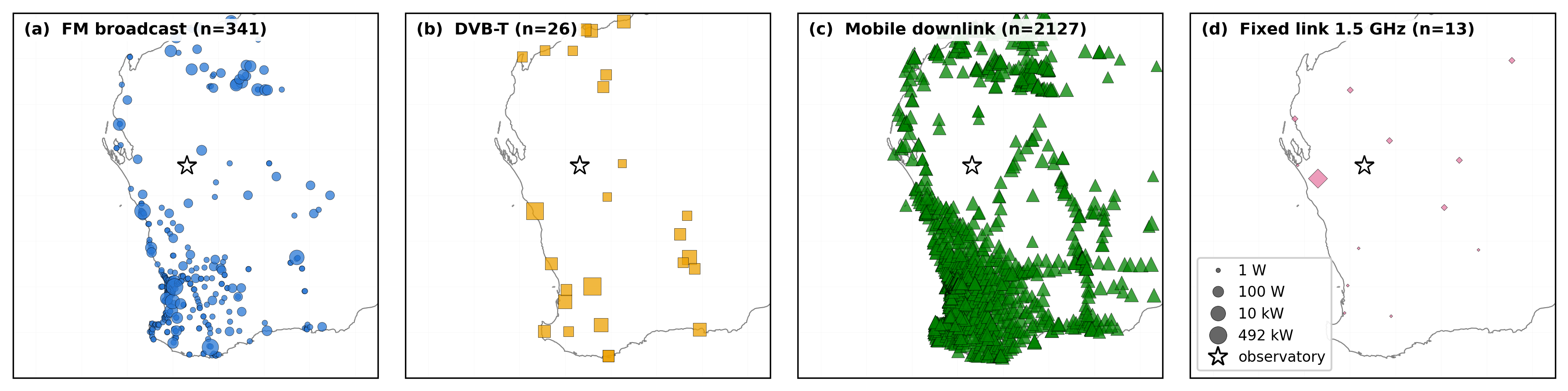}
\caption{Attributed transmitter sites in the Australia-west region from the 2023 licence snapshot, by service group, over all 23 licensed channels. Marker area scales with licensed \ac{EIRP} from 1 W to 492 kW, clipped at that value; the star marks the observatory.}
\label{fig:txmap}
\end{figure*}

The 2023 snapshot yields 9\,479 licensed transmitter-channel entries across the monitored channels, de-duplicated to 9\,404 unique channel-sites. Per-channel populations range from 6 (536 MHz) to 1\,281 (778 MHz). Licensed \acp{EIRP} reach 492 kW for the 631 MHz \ac{DVB-T} channel. Seven records matched to the 1865 MHz channel carry a nominal 10 MW \ac{EIRP} against a stated transmit power of 7.94 W, an implied antenna gain of 61 dB that is not credible for the lightpole small cells they describe. All seven belong to a single licence, and all are emissions centred at 1857.5 MHz whose 15 MHz bandwidth reaches the channel edge. We retain them rather than editing the register. The excess predictor of Section~\ref{sec:pathloss} is a difference of two losses along the same path and does not depend on \ac{EIRP}, so the analytical results are unaffected; these records do, however, win the absolute received-power feature on 76\% of hours at 1865 MHz, and that channel carries no fitted weights in the forecast product (Section~\ref{sec:calibration}). Path lengths to the site span 98 to 1\,005 km; transmitters outside the Australia-west region of Section~\ref{sec:gfs} are excluded. The register misses spectrum-licensed deployments, which is exactly why the \ac{IMT} blocks at 2.6 GHz required the empirical same-tower test of Section~\ref{sec:imt}. Figure~\ref{fig:txmap} maps the population by service group.

\section{Basic transmission loss modelling}
\label{sec:pathloss}

For every transmitter-channel path and every hourly timestep we compute the terrain-aware basic transmission loss $L_b$ to the site, in the sense of Recommendation ITU-R P.341 \citep{itu_p341}. Terrain profiles are sampled at 90 m steps and cached per site; production $L_b$ uses the Copernicus GLO-90 digital elevation model \citep{copernicus_dem}, replacing the SRTM3 data \citep{farr2007} used initially after a void defect was discovered (Section~\ref{sec:terrain}). The receiver height is 21 m, matching the \ac{RFI} tower antenna.

Channels at 100 MHz and above use Recommendation ITU-R P.452-16 \citep{itu_p452_16} as implemented in pycraf 2.1.0 \citep{pycraf_ascl, winkel2018}. P.452 treats diffraction, tropospheric scatter, and crucially, ducting and elevated-layer reflection, but its stated validity begins at about 100 MHz. The \ac{FM} channels below that use Recommendation ITU-R P.1812-6 \citep{itu_p1812_6}, a path-specific point-to-area method valid from 30 MHz, via the Py1812 6.1 implementation \citep{py1812}, which is validated against the ITU-R reference test profiles. P.1812 restricts the time percentage $p$ to the range 1--50\%.

Both recommendations have since been revised. Between P.452-16 and P.452-18 the ducting and layer-reflection model is unchanged, troposcatter is formula-identical with only a change in how the sea-level refractivity map is delivered, and the clutter model was removed to a separate recommendation, it is not used here; for the paths in this work the two revisions are therefore numerically equivalent. Between P.1812-6 and P.1812-7 the ducting model is likewise unchanged, but troposcatter was replaced by a physically-based formulation with explicit dependence on the refractivity gradient. P.1812 is used only for the \ac{FM} channels, where troposcatter sets the background level on long trans-horizon paths, so that revision is a real change for those channels; it does not affect the results because those channels return nulls, but a future positive \ac{FM} result should be recomputed under P.1812-7 first.

Both are statistical models: they predict the loss not exceeded for a percentage $p$ of an average year, parameterised by the sea-level refractivity $N_{0}$ and the refractivity lapse $\Delta N$ over the lowest 1 km. For each timestep, we evaluate $\Delta N$ and $N_{0}$ from the \ac{GFS} columns along each great-circle path and take the path median of $\Delta N$. The predicted excess power is the standard-atmosphere reference loss minus the per-timestep loss. This hybrid use of a statistical model as an instantaneous one makes $p$ an effective calibration constant; Section~\ref{sec:validation} sets it by validation. The approach is sensitive where it needs to be: on a 956 MHz Geraldton-to-site test path, the P.452 ducting term swings $\sim$75 dB between standard and strongly-ducting parameterisations. At 88 MHz the P.1812 loss spans only $\sim$25 dB across $\Delta N$ values of 30 to 150 N-units per km, reflecting the weaker duct response at \ac{VHF}.

That span is itself a function of $p$. In the recommendations $p$ is the percentage of an average year for which the predicted loss is not exceeded, and it indexes the anomalous-propagation mechanisms: a small $p$ selects the rare, most strongly enhanced conditions, a large $p$ the typical ones. We do not assert that any given hour's loss is exceeded for $p$ of the year. We hold $p$ fixed and let all of the hour-to-hour variation enter through $\Delta N$ and $N_{0}$ along the path, so $p$ ceases to be a probability and becomes a regime selector: it fixes how much anomalous propagation the model is willing to assume, and the instantaneous refractivity gradient then modulates the loss within that regime. Because the predictor is scored by rank its absolute level is irrelevant, and only the ordering matters. The useful $p$ is therefore the one that maximises the swing quoted above, since that swing is what separates ducted hours from quiet ones.

\section{Validation}
\label{sec:validation}

\subsection{The skill metric}

We score with a stratified \ac{AUC} throughout. A pooled \ac{AUC} credits a channel for the seasonal and diurnal cycle it shares with the predictor. Measured against a predictor given only month and hour of day and no meteorology at all, 12 of the 25 channels have a pooled \ac{AUC} below that baseline. Reporting pooled numbers would overstate the low-frequency channels badly.

The stratified statistic, a van Elteren test, compares event hours only against non-event hours in the same month and the same three-hour block, then pools strata weighted by their discordant-pair counts. Every comparison is therefore between two hours at the same time of day in the same season. This can be validated: on the same span and channels the number passing an \ac{AUC} $\geq 0.6$ gate rises from 17 to 20 of 25, and no channel is lost. The 25-channel median is 0.724 stratified against 0.726 pooled, but the ranking is reordered, which is the point.

Every channel carries a 95\% confidence interval from a block bootstrap on 7-day blocks; resampling by hour would badly understate the interval because ducting events persist for hours. Pooled and stratified numbers are never mixed between sections.

\subsection{Evaluation design}
We evaluate with three deliberately different train/test designs: a frozen-transfer test that calibrates the time percentage on the pilot months and then applies it unchanged to the years the optimiser never saw; an interleaved year-block split that re-selects the time percentage on alternate years so \ac{ENSO} phases enter both sets, reducing our sensitivity to climate-phase artefacts; and a contiguous future-holdout for the learned model of Section~\ref{sec:discussion}. In short: the physical calibration constant is estimated on a climate-balanced split, the deployable learned models are tested only on data from after their training period, and the main result uses a calibration fixed once on the pilot and never adjusted again.

A fourth division is imposed by the data, and we use it as an additional test. \ac{GFS} changed model version on 2021 March 22. We train and score on the v16 era only and hold the v15 era out entirely, so every channel carries a validation column from a genuinely different \ac{NWP} model version that the weights never saw. This was decided on the strength of the claim rather than on the numbers: including v15 in training is worth a median 0.004 \ac{AUC} and wins on 7 of 25 channels while losing significantly on none. Giving up 0.004 \ac{AUC} to validate a cross-model-version is a good trade.

\subsection{Operating point and per-channel skill}
\begin{figure}[t]
\includegraphics[width=\columnwidth]{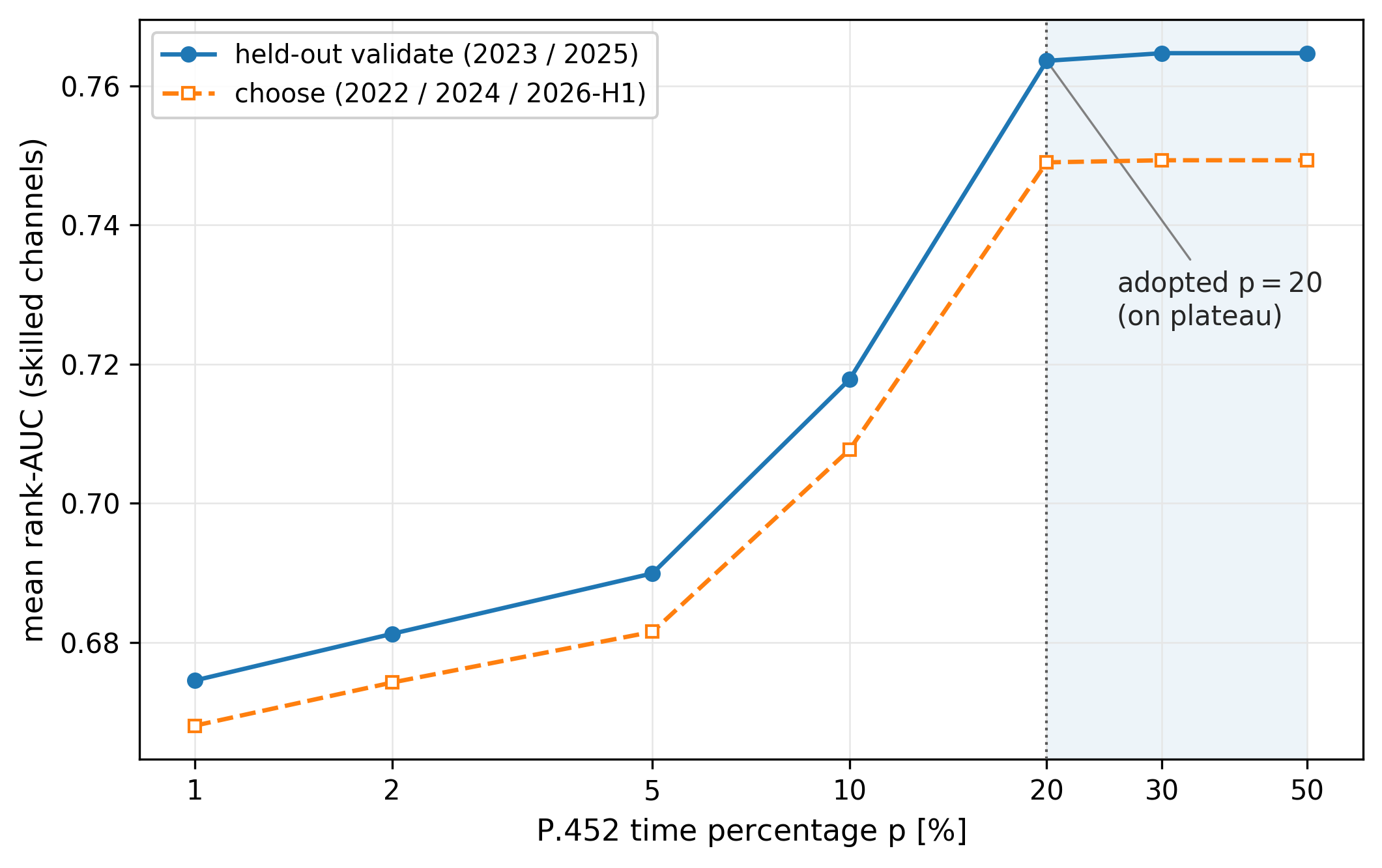}
\caption{Selecting the P.452 time percentage $p$. Mean rank-AUC over the skilled channels versus $p$ on the interleaved year-block split: $p$ chosen on 2022, 2024, and 2026 (dashed), confirmed on the held-out 2023 and 2025 (solid). Skill saturates at $p=20\%$ and stays flat to $p=50\%$ (shaded); $p=20\%$ is adopted as the lowest, most conservative value at full skill, coincident with the pilot calibration.}
\label{fig:poptim}
\end{figure}

We sweep the time percentage $p$ over $\{1, 2, 5, 10, 20, 30, 50\}$ at full resolution on the interleaved year-block split. Skill rises and then saturates: on the held-out years it is flat from $p=20\%$ to $p=50\%$ (0.764, 0.765, 0.765), with a lower tail of 0.674 at $p=1\%$. We adopt $p=20\%$, the lowest value on the plateau and so the most conservative footprint at full skill (Figure~\ref{fig:poptim}). That choice, made on the full archive, coincides with the independent pilot calibration, verifying the frozen-transfer result sits on a broad plateau.

The shape of that curve is the opposite of the natural expectation. P.452 scales the effective earth radius by $k_{50} = 157/(157 - \Delta N)$, which is singular at $\Delta N = 157$ N-units per km, precisely the super-refraction threshold at which $\mathrm{d}M/\mathrm{d}z$ reaches zero: the recommendation is undefined inside a duct, and $\Delta N$ can only ever approach ducting from below. The consequence of this is that sweeping $\Delta N$ from 30 to 150 N-units per km on a smooth 420 km path at 956 MHz, the spread in predicted total loss grows from 10.3 dB at $p=0.1\%$ to 18.7 dB at $p=1\%$, 25.3 dB at $p=10\%$ and 26.1 dB at $p=20\%$, and is then flat. At small $p$ the ducting term already sits near the most favourable value it admits, so the loss stops responding to the gradient, and ducted hours compress towards quiet ones. The skill sweep follows this almost exactly: across the swept values the rank correlation between the model's dynamic range and the held-out \ac{AUC} is 0.97. Selecting $p$ is therefore less the tuning of a free parameter against labels than the anchoring of the model in the regime where its duct response is steepest, with the labels confirming it independently.

\begin{figure*}
\includegraphics[width=\textwidth]{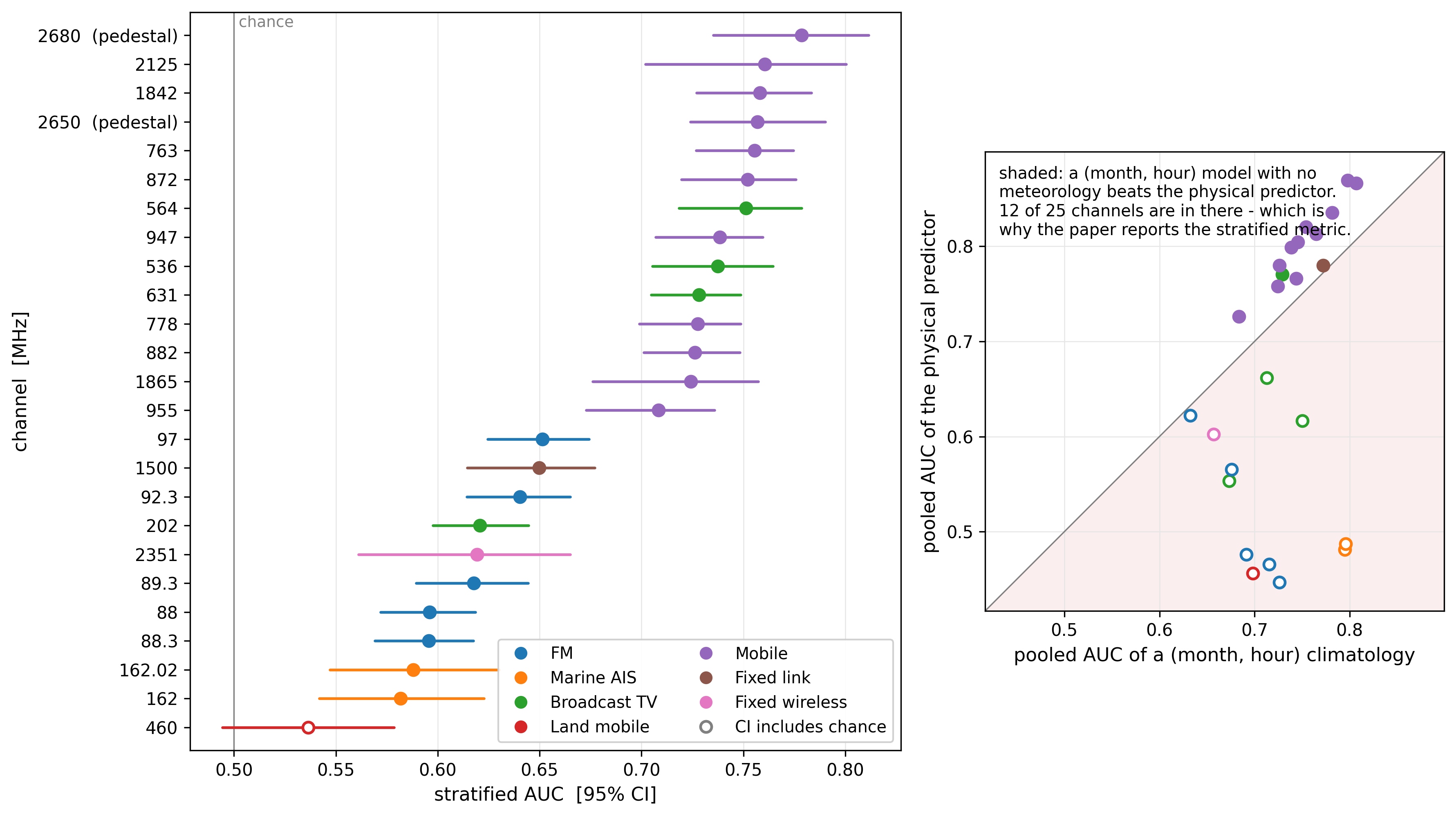}
\caption{Left: stratified \ac{AUC} per channel with 95\% block-bootstrap intervals, coloured by service group; 460 MHz is unfilled and straddles chance. Right: pooled \ac{AUC} against each channel's own pooled climatology, with the region in which climatology wins shaded. Twelve of the 25 channels fall in that region, which is the argument for scoring within strata.}
\label{fig:channelskill}
\end{figure*}

\begin{figure*}
\includegraphics[width=\textwidth]{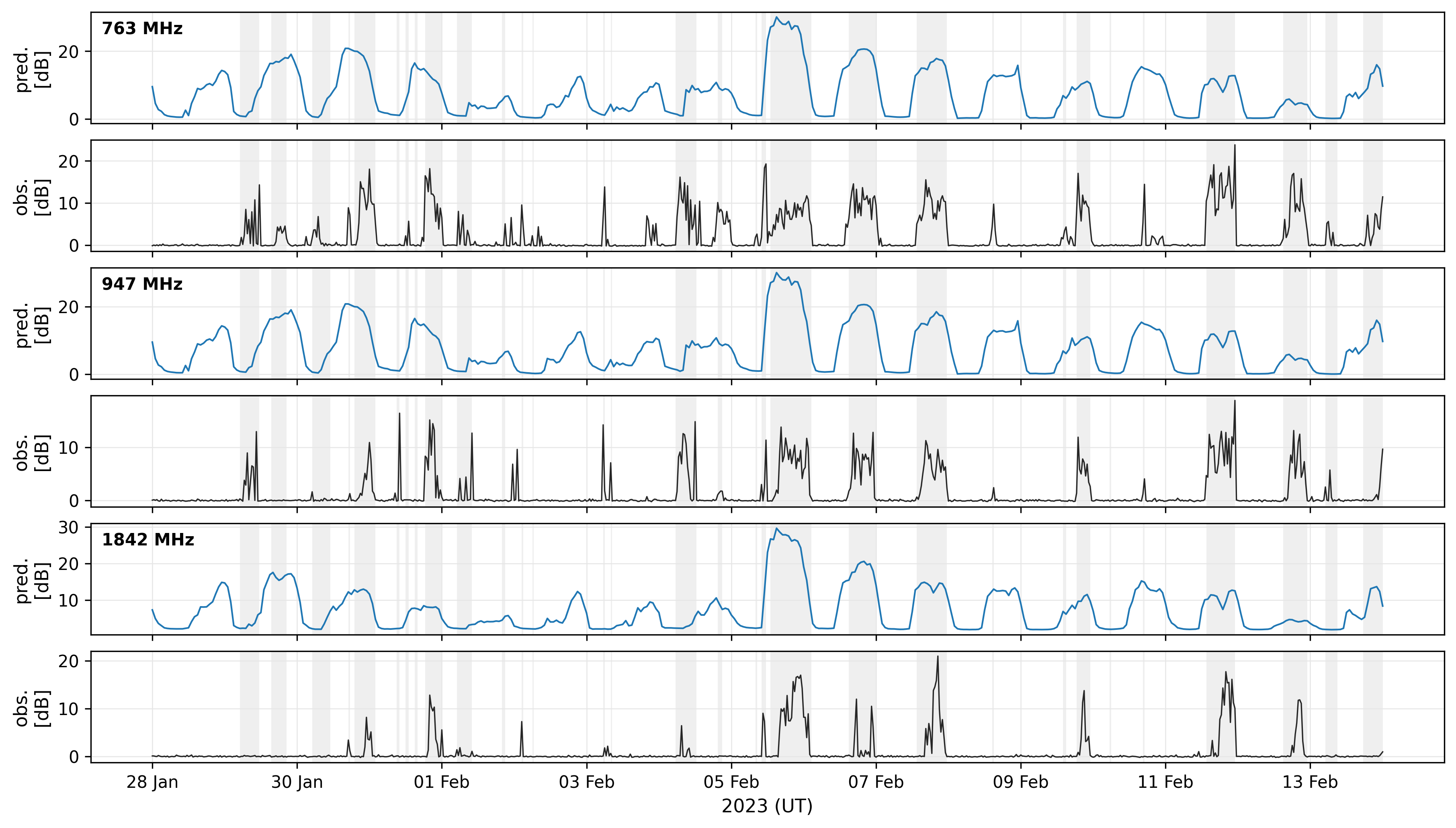}
\caption{Analytical predicted excess power (blue) and observed received power above the rolling 7-day median baseline (black) for the 763, 947, and 1842 MHz channels, 2023 January 28 to February 14. Shading marks flagged multi-channel event periods. This is the pure P.452 physics prediction from the continuous refractivity profiles, upstream of the calibrated duct decision and the learned models of Section~\ref{sec:discussion}. The predicted panel is drawn from an earlier reduction and is illustrative; it contributes no number to Table~\ref{tab:skill}.}
\label{fig:predobs}
\end{figure*}

Table~\ref{tab:skill} and Figure~\ref{fig:channelskill} give the per-channel result. The 25-channel median stratified \ac{AUC} is 0.724, and 24 of the 25 confidence intervals exclude chance. The strongest channels are 2680 (0.779), 2125 (0.761), 1842 (0.758), 2650 (0.757), 763 (0.756), 872 (0.752) and 564 (0.751) MHz. At the bottom, 460 MHz reads 0.536 with an interval of 0.494 to 0.579 and is not distinguishable from chance; the \ac{AIS} and \ac{FM} channels cluster between 0.58 and 0.66. The v15 column, the same analytic predictor evaluated out of era on a different model version, tracks the v16 column closely on the well-populated channels. Figure~\ref{fig:predobs} shows an eventful pilot fortnight: the predicted excess rises through the flagged event periods, including the strong February 5--6 2023 episode mapped in Figure~\ref{fig:ductcase}.

\begin{table*}[t]
\caption{Per-channel skill in the production configuration ($p=20\%$, 21 m receiver height, GLO-90 terrain, surveyed site position). $N_{\mathrm{tx}}$ is the attributed transmitter count and $N_{\mathrm{ev}}$ the event hours in the analysis era. \ac{AUC} is the stratified (van Elteren) statistic, scored within month and three-hour strata so that the seasonal and diurnal cycle shared with the predictor cannot inflate it; the 95\% interval is a block bootstrap on 7-day blocks. \ac{AUC}$_{\mathrm{v15}}$ is the same statistic on the held-out \ac{GFS} v15 era, a different model version; that era spans 5\,274 hours, so several of its intervals are wide. \ac{BSS} is the Brier skill score of the operational probability against each channel's own climatology, scored out of sample against weights fitted on the training period alone (Section~\ref{sec:calibration}); it is blank where the channel is published without fitted weights. $^{\dagger}$The interval includes 0.5: not distinguishable from chance.}
\label{tab:skill}
{\tablefont\begin{tabular}{@{\extracolsep{\fill}}rlrrccc}
\toprule
$f$ (MHz) & Service & $N_\mathrm{tx}$ & $N_\mathrm{ev}$ & AUC$_\mathrm{v16}$ [95\% CI] & AUC$_\mathrm{v15}$ & BSS \\
\hline
88 & FM & 314 & 3419 & 0.596\,[0.572,0.619] & 0.594\,[0.54,0.65] & +0.059 \\
88.3 & FM & 225 & 3333 & 0.596\,[0.569,0.617] & 0.585\,[0.54,0.64] & +0.074 \\
89.3 & FM & 20 & 3322 & 0.618\,[0.589,0.644] & 0.579\,[0.52,0.63] & +0.080 \\
92.3 & FM & 34 & 3348 & 0.640\,[0.614,0.665] & 0.646\,[0.60,0.69] & +0.048 \\
97 & FM & 33 & 2531 & 0.651\,[0.624,0.674] & 0.663\,[0.60,0.71] & -- \\
162 & Marine AIS & 184 & 2761 & 0.582\,[0.542,0.623] & 0.590\,[0.54,0.64] & +0.144 \\
162.02 & Marine AIS & 184 & 2686 & 0.588\,[0.547,0.630] & 0.589\,[0.54,0.64] & +0.143 \\
202 & DVB-T & 10 & 4589 & 0.621\,[0.598,0.645] & 0.622\,[0.57,0.67] & +0.045 \\
460 & Land mobile & 298 & 3831 & 0.536\,[0.494,0.579]\textsuperscript{$\dagger$} & 0.609\,[0.52,0.68] & +0.031 \\
536 & DVB-T & 6 & 2121 & 0.737\,[0.705,0.765] & 0.705\,[0.64,0.76] & +0.064 \\
564 & DVB-T & 9 & 1955 & 0.751\,[0.718,0.779] & 0.694\,[0.64,0.74] & +0.071 \\
631 & DVB-T & 16 & 5283 & 0.728\,[0.705,0.749] & 0.718\,[0.67,0.76] & +0.184 \\
763 & Mobile & 787 & 5429 & 0.756\,[0.727,0.775] & 0.730\,[0.68,0.77] & +0.230 \\
778 & Mobile & 1281 & 6878 & 0.728\,[0.699,0.749] & 0.723\,[0.68,0.77] & +0.246 \\
872 & Mobile & 551 & 2274 & 0.752\,[0.720,0.776] & 0.706\,[0.64,0.77] & -- \\
882 & Mobile & 1209 & 5388 & 0.726\,[0.701,0.748] & 0.721\,[0.68,0.77] & +0.106 \\
947 & Mobile & 863 & 3355 & 0.738\,[0.707,0.760] & 0.727\,[0.67,0.78] & +0.178 \\
955 & Mobile & 564 & 2200 & 0.708\,[0.673,0.736] & 0.714\,[0.64,0.77] & +0.102 \\
1500 & Fixed link & 13 & 2559 & 0.650\,[0.614,0.677] & 0.678\,[0.63,0.72] & +0.118 \\
1842 & Mobile & 459 & 1484 & 0.758\,[0.727,0.783] & 0.728\,[0.63,0.80] & +0.199 \\
1865 & Private LTE & 992 & 754 & 0.724\,[0.676,0.757] & 0.697\,[0.56,0.79] & -- \\
2125 & Private LTE & 484 & 389 & 0.761\,[0.702,0.800] & 0.692\,[0.58,0.77] & -- \\
2351 & Fixed wireless & 586 & 281 & 0.619\,[0.561,0.665] & 0.530\,[0.37,0.67] & -- \\
2650 & Mobile (IMT) & 541 & 1299 & 0.757\,[0.724,0.790] & 0.786\,[0.71,0.85] & +0.194 \\
2680 & Mobile (IMT) & 541 & 409 & 0.779\,[0.735,0.812] & 0.778\,[0.69,0.84] & -- \\
\botrule
\end{tabular}}

\end{table*}

\subsection{Controls}
Three checks argue that the skill reflects atmospheric information rather than seasonal or diurnal artefacts. The first is the stratification itself: scoring within month and three-hour strata removes the shared cycle, which is why a pooled statistic credits 12 of the 25 channels with a seasonal and diurnal cycle they share with the predictor, and loses to a month-and-hour model on exactly those channels. Second, a transmitter-agnostic baseline predictor, the per-hour fraction of land grid cells more than 100 km from the site whose columns duct (minimum $\mathrm{d}M/\mathrm{d}z < 0$ below 3 km), reaches \ac{AUC} 0.63--0.72 on the duct-driven channels in the summer-rich pilot window and 0.53--0.69 over the full record, on the same nine channels: coarse duct occurrence alone carries much of the signal, and the per-path modelling adds the remainder. Third, the skill is robust to the underlying model, as follows.

Running the identical pipeline with different atmospheric models tests what input fidelity can (or cannot) improve. A 137-level ERA5 reconstruction \citep{hersbach2020} reproduces the \ac{GFS} duct climatology, yet through the ITU coupling \ac{GFS} beats ERA5 by 0.02--0.04 on every skilled channel: the finer vertical grid does not help, while ERA5's coarser $\sim$31 km horizontal grid (\ac{GFS}: $\sim$13 km native) results in an inferior match.
The 4.4 km BARRA-C2 regional reanalysis \citep{su2025} completes the three-model resolution trial. On discrimination by absolute received power the trial ladder is monotone (mean \ac{AUC} 0.667, 0.683, 0.704 for ERA5, \ac{GFS}, and BARRA-C2, with BARRA-C2 beating \ac{GFS} on 9 of 10 validatable channels); on the excess metric it saturates (0.690, 0.719, 0.715). The step from 31 to 13 km improves skill everywhere; the step from 13 to 4.4 km only pays for absolute-power discrimination, plausibly because the path-median $\Delta N$ coupling smooths away the narrow duct corridors that BARRA-C2 resolves (taken up in Section~\ref{sec:corridor}). An independent evaluation confirmed the comparison on identical event populations and equivalent code paths.

\begin{figure*}
\includegraphics[width=\textwidth]{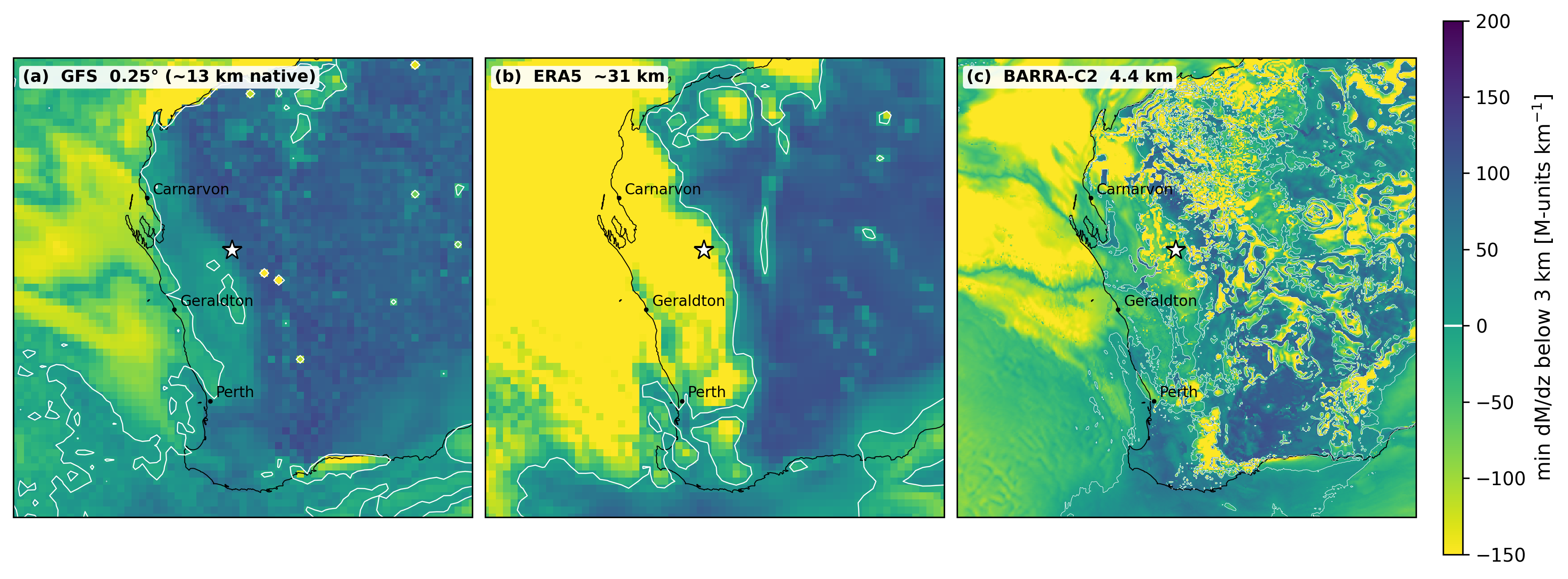}
\caption{Minimum vertical gradient of modified refractivity below 3 km at the 2023 February 5 21:00 UT case-study hour, from the three driving models at their native resolutions: (a) \ac{GFS}, (b) ERA5, (c) BARRA-C2. Fields are shown at the physical duct onset, before the calibrated duct decision of Section~\ref{sec:corridor} is applied. The white contour marks $\mathrm{d}M/\mathrm{d}z = 0$, the duct boundary; the star marks the observatory. Horizontal resolution controls how much duct structure the input resolves.}
\label{fig:modelcmp}
\end{figure*}

\begin{figure*}
\includegraphics[width=\textwidth]{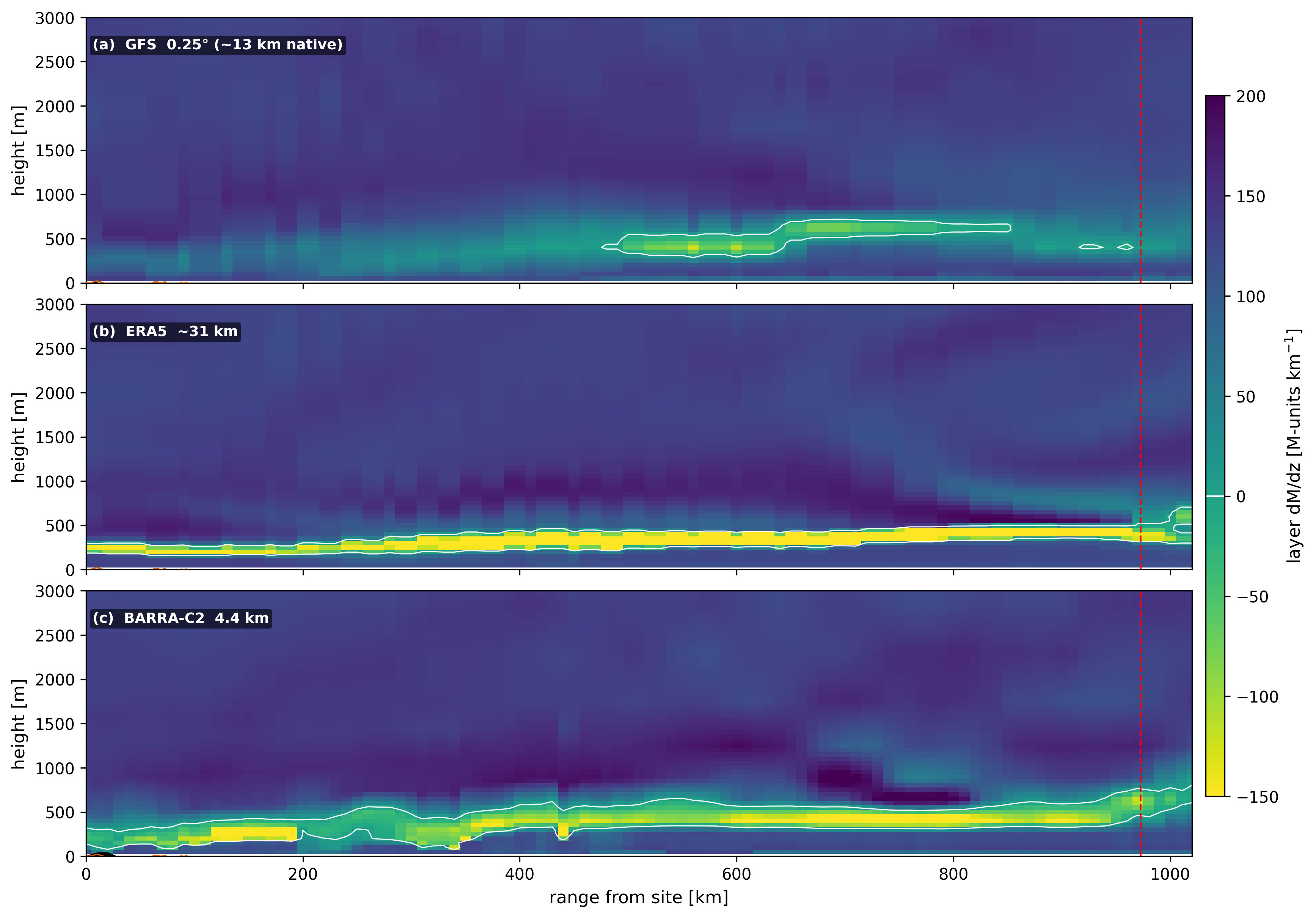}
\caption{Range-height sections of the layer gradient $\mathrm{d}M/\mathrm{d}z$ along the great-circle path from the observatory to the most distant decoded \ac{AIS} ship (972 km, azimuth 316$^{\circ}$) at 2023 February 5 21:00 UT, from (a) \ac{GFS}, (b) ERA5, and (c) BARRA-C2. The white contour marks $\mathrm{d}M/\mathrm{d}z = 0$; the dashed line marks the ship's range. All three models produce an elevated duct below $\sim$700 m bridging ship to site, but its continuity and depth differ between models at this hour.}
\label{fig:toiprofiles}
\end{figure*}

Figures~\ref{fig:modelcmp} and \ref{fig:toiprofiles} make the resolution contrast concrete at the case-study hour of Figure~\ref{fig:ductcase}. The duct-field maps sharpen from 31 through 13 to 4.4 km, and the range-height sections along the 972 km path to the most distant decoded \ac{AIS} ship show an elevated duct below $\sim$700 m that ERA5 and BARRA-C2 carry continuously from ship to site, while \ac{GFS} resolves it in only about 40\% of the path columns. Instantaneous structure does not rank like the statistical ladder; the ladder is a statement about time-aggregated skill, not per-timestep fidelity.

\subsection{Forecast lead time}
\begin{figure*}
\includegraphics[width=\textwidth]{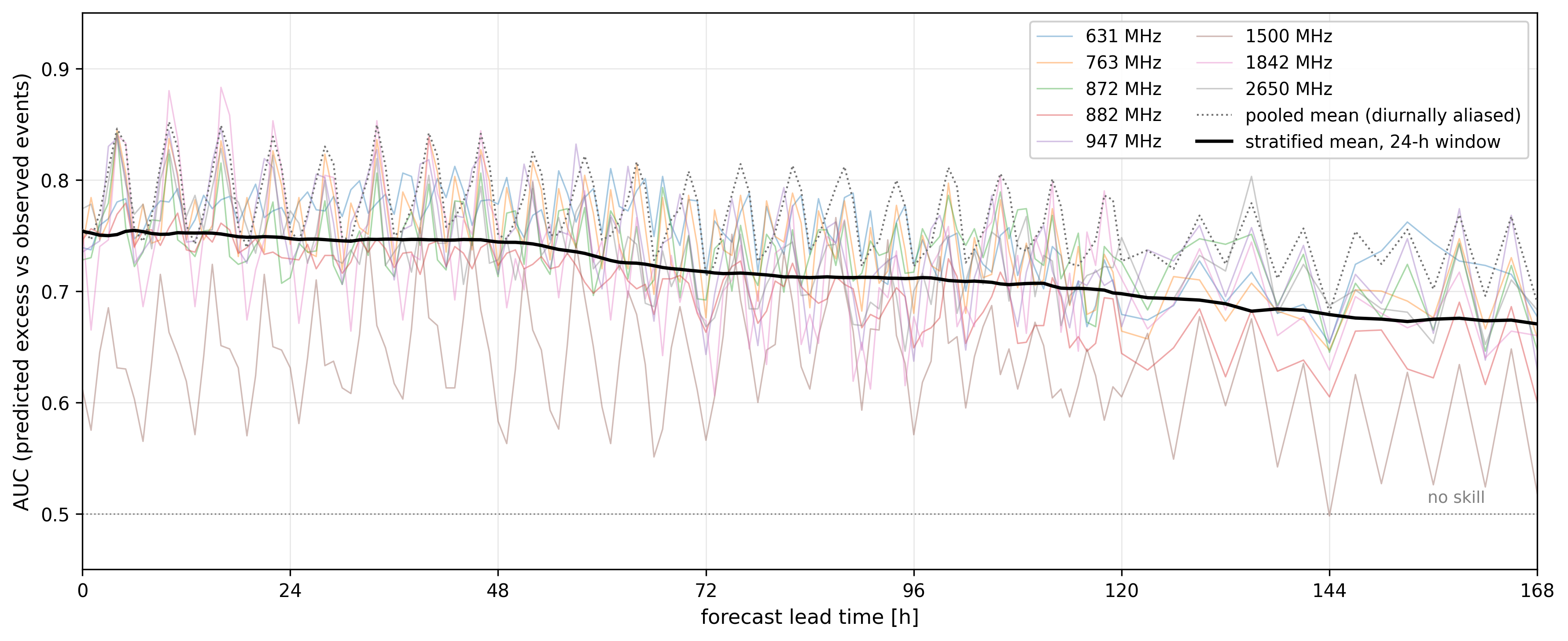}
\caption{Stratified \ac{AUC} versus \ac{GFS} forecast lead time (f000--f168) for the skilled channels (thin lines), scored over ${\sim}110$ forecast days in February and June windows of 2022--2025. The bold line is the stratified mean over a centred 24-h lead window; the dotted line is the pooled mean, which is diurnally aliased: a fixed lead verifies on only four clock hours, so hour-set composition drives its 6-hour zigzag and props up its long-lead level.}
\label{fig:leadskill}
\end{figure*}

Forecast lead time erodes the skill only gradually, but measuring this requires care (Figure~\ref{fig:leadskill}). The archive holds four daily forecast cycles, so a fixed lead verifies on only four clock hours, and because ducting is strongly diurnal a pooled \ac{AUC} shows the diurnal climatology itself: the pooled curve zigzags with a 6-hour period and its long-lead level is propped up by composition rather than skill. Stratifying the \ac{AUC} within the valid clock hour removes the artefact. Scoring \ac{GFS} forecast hours f000--f168, the stratified mean over the skilled channels reads 0.74 through three days of lead, 0.71 over days four and five, 0.69 on day six and 0.67 on day seven: flat through day three, then a step. The day-one-to-three plateau is reproduced in February windows of 2024 and 2025 within $\pm$0.03 of the 2023 value; two-week winter windows are too event-starved to score, so the lead-time result is a statement about the event season, which is where the operational value lies. The verification windows are event-rich at every lead: the mean event count per lead stays above 34 event-hours across f000--f168, so the decay is not a small-sample artefact, though individual sparse channels do grow noisy at long lead.

\section{Discussion}
\label{sec:discussion}

\subsection{Mobile bands: positive attribution}
The site \ac{LTE} cell scanner independently confirms both the events and their sources. During ducting conditions it decodes over-the-horizon cells; 86--97\% of its detections fall within flagged event hours, the range being across the four monitored \ac{LTE} channels rather than an uncertainty. The decoded SIB1 network identities match the \ac{ACMA} operator attribution per channel: Optus at 763 MHz, Telstra at 778 MHz, Vodafone at 872 MHz, and Optus at 1842 MHz. Cell-level geolocation through a cached cell database locates 95--100\% of decodes, and every decoded cell has an \ac{ACMA} licensed site within 30 km. Ranking the co-channel candidates by absolute EIRP-weighted received level places the true source at a median rank of 4 among $\sim$10$^{3}$ candidates, with top-5 hit rates of 62--87\%. Ranking by ducting excess instead fails (top-5 hit rates 8--25\%): during an event the duct enhances hundreds of paths at once, so the discriminating quantity is the absolute level, not the enhancement. This is a methodological trap for source identification under ducting conditions. Decoded sources include Mullewa, Denham, and Carnarvon at 763 MHz, Carnarvon at 872 MHz (a single cell accounts for 84\% of decodes), and Exmouth at 589 km at 1842 MHz. The scanner is duty-cycled and positive-only: detections are confirmations, but their absence proves nothing.

\subsection{VHF ground truth from AIS ship receptions}
\label{sec:ais}
The site \ac{AIS} decoder supplies an independent \ac{VHF} validation set at 162 MHz. Every \ac{AIS} message carries the transmitting ship's own GNSS position, so each reception is a self-located $\sim$12 W transmitter at a known position at sea. The archive holds $\sim$470\,000 position reports from 2\,418 vessels between 2018 February and 2024 November. Search-and-rescue aircraft (five \ac{AMSA} airframes, 67\,000 messages) are excluded by filtering \ac{AIS} message type 9 and the \ac{MMSI} prefix 111. Reception distances are far beyond the radio horizon: median 485 km, 95th percentile 914 km, maximum 2\,357 km.

The 1\,572 receptions within the pilot period test the duct diagnostics directly. Relative to matched control samples on the same paths within $\pm$14 days, reception paths show a land-segment duct fraction of 0.259 versus 0.155 (67\% higher), a maximum duct depth along the path of 28.1 versus 18.8 M-units (49\% higher), and a contiguous duct bridging the coastline on 87.4\% versus 75.4\% of paths. The sea segments are near-saturated with ducting in season (0.979 versus 0.938): what gates reception is the marine duct extending overland to the site. Like the \ac{LTE} scanner, the \ac{AIS} logger is duty-cycled and unreliable (and stale since 2024 November), so it provides positive evidence only. The result demonstrates that \ac{VHF} ducting into the site occurs and is captured by the GFS-derived diagnostics.

\begin{figure*}
\includegraphics[width=\textwidth]{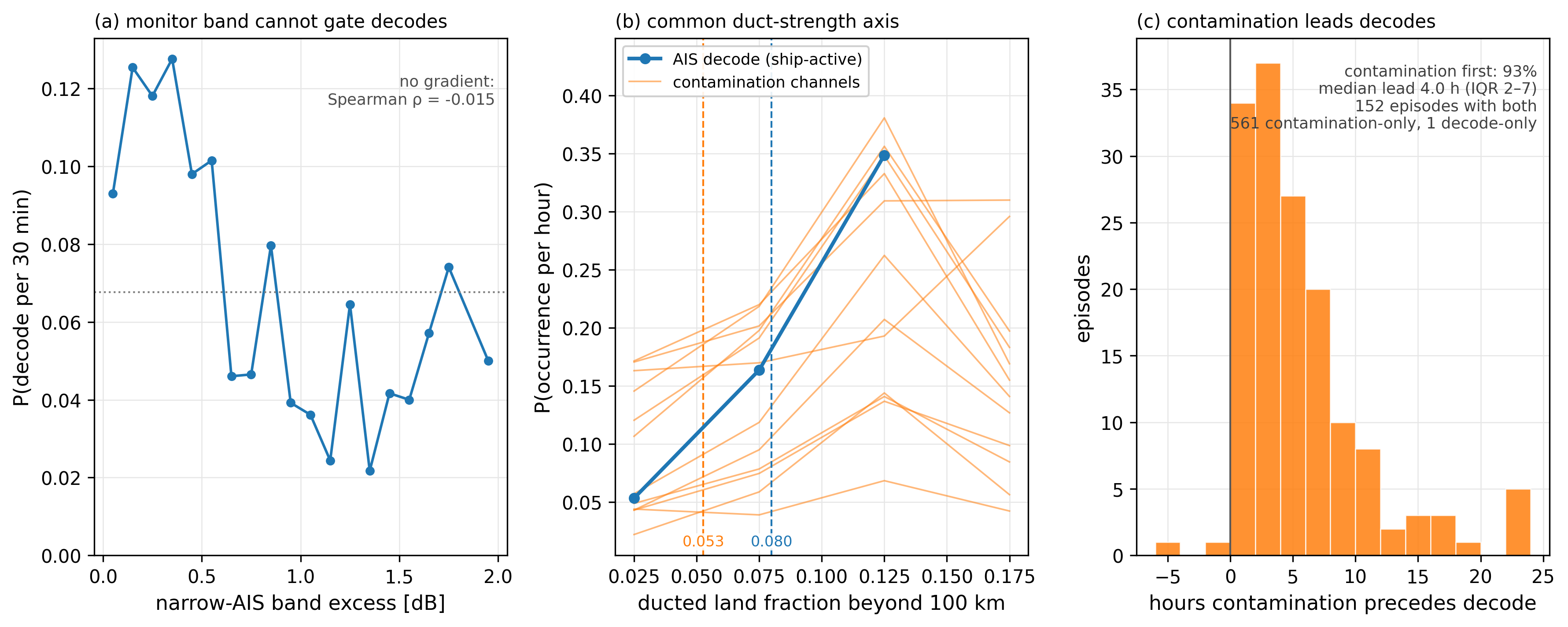}
\caption{AIS decodes as a ducting marker. (a) decode probability against same-frequency band excess, showing no positive gradient (Spearman $\rho=-0.015$), because ship transmissions arrive below the swept monitor's displayed floor. (b) decode and contamination hours on a common duct-strength axis, where decodes sit at higher duct strength. (c) distribution of the lead of telescope contamination over the first decode within an episode.}
\label{fig:aismarker}
\end{figure*}

The decoder and the swept monitor differ significantly in sensitivity. A 12.5 W Class A AIS transmitter at these ranges arrives at 4 to 18 dB$\mu$V. The monitor's displayed floor in this band is about 35 dB$\mu$V; the decoder works down to about 0 dB$\mu$V. Every reception here is therefore 15 to 30 dB below what the monitor can register. Decode probability is accordingly flat against the 162 MHz band excess (Figure~\ref{fig:aismarker}a): the band carries no information about ship reception. Because of that gap, the comparison must be made on duct strength, not in monitor-band units.

On duct strength the two populations separate (Figure~\ref{fig:aismarker}b). Decodes occur at a median ducted land fraction of 0.080. Contamination-event hours sit at 0.053, spanning 0.045 to 0.065 across the twelve channels, against an all-hours background of 0.055. A decode therefore needs a stronger duct than contamination does. The episode counts show this: of 714 episodes, 561 contaminate with no decode, 152 show both, and one shows a decode alone. Where both occur, contamination comes first in 93\% of episodes, by a median of 4.0 hours (Figure~\ref{fig:aismarker}c). Decodes are a specific but insensitive marker. A decode confirms a well-developed duct; their absence proves nothing, and they cannot bound the onset of contamination. The decode side is further conditioned on the presence of ships with AIS transmitters, so absolute rates are not comparable between the two populations; only the duct-strength positions and the episode ordering are.

\subsection{FM channels: a different mechanism}
The \ac{FM} channels carry far less ducting skill than the \ac{UHF} and microwave channels, and what remains is too weak to support a ducting mechanism. This holds for P.1812, for a P.452 run clamped to its frequency floor, and for every raw duct diagnostic. The two propagation models agree with $r\!\approx\!1.0$, so the result is not an artefact of either.

Over 3\,419 event hours 88 MHz reads a stratified \ac{AUC} of 0.596, with an interval of 0.572 to 0.619; 88.3 MHz reads 0.596 (0.569 to 0.617) and 89.3 MHz 0.618 (0.589 to 0.644). These intervals exclude chance, so an accurate description is weak skill rather than a null: the only channel of the 25 whose interval contains 0.5 is 460 MHz. But the contrast with the duct-driven channels, which run from 0.73 to 0.78, is a factor of several in the separation from chance, and weak residual skill is what the near-threshold mechanism proposed below would itself produce, since reception at the margin is modulated by ordinary refractive variability that the duct predictor partly tracks. 97 MHz sits a little higher again at 0.651 (0.624 to 0.674). An earlier assessment called that channel inconclusive; a contaminated reference band caused that reading, and it is superseded.

Two alternative long-range mechanisms are excluded. The first is sporadic-E. Autoscaled foEs from the Learmonth ionosonde never exceeded 18.8 MHz during the pilot. Single-hop Es therefore reaches about 94 MHz at best, and only 0.2\% of 88 MHz event hours are geometrically feasible. foEs itself scores 0.5 to 0.6 against every monitored channel. Upper-air ionospheric state is irrelevant to these events.

The second is aircraft scatter. Hourly counts of aircraft above 20\,000 ft in the corridor to the coastal \ac{FM} transmitters correlate weakly with \ac{FM} events, at \ac{AUC} 0.56. That is diurnal confounding: flights and \ac{FM} events are both daytime. Stratifying by hour of day collapses it to 0.48--0.55. The ducting channels show the mirror image, 0.41 raw from nocturnal anticorrelation and about 0.52 stratified. That contrast is a useful internal control.

Near-threshold reception of local sources remains. The Jack Hills mine site lies 94 km away and carries 49 W \ac{EIRP} on both monitored \ac{FM} frequencies. Low-power open narrowcasting stations lie at 143--300 km. The \ac{FM} events follow a flat daytime pattern, which fits threshold crossing by nearby weak transmitters under ordinary refractive variability. Section~\ref{sec:ais} shows that ducting at comparable \ac{VHF} frequencies is real and well captured by the diagnostics. The \ac{FM} result is therefore about attribution, not about physics.

\subsection{Null and negative results}
\label{sec:nulls}
Beyond the \ac{FM} attribution, several negative results deserve prominence, because each contradicts a plausible default assumption.
\begin{itemize}
\item 460 MHz is not distinguishable from chance. Its interval, 0.494 to 0.579 on 3\,831 event hours, is the only one of the 25 that includes 0.5. It is a well-powered null, not a weakly skilled channel.
\item Higher vertical resolution does not improve fidelity. The 137-level ERA5 reconstruction reaches diagnostic parity with the 21-level \ac{GFS} profiles, then loses to \ac{GFS} through the ITU coupling on every skilled channel. The path coupling rewards horizontal resolution, not vertical.
\item No transmitter-free diagnostic approaches the ITU-coupled skill. An initial BARRA-C2 result of 0.706 did not survive audit: thin 20--40 m cross-source layers dominated it. With a 40 m thickness floor the transmitter-free diagnostics score 0.51--0.66 across the three models.
\item Ranking interference candidates by ducting excess fails for source identification. Top-5 hit rates are 8--25\%, against 62--87\% for absolute EIRP-weighted level.
\item Small time percentages degrade skill. At $p = 0.1$--1\% the ducting term already sits near its most favourable value, so the predicted loss stops responding to the refractivity gradient and the predictor loses more than half its dynamic range (Section~\ref{sec:validation}). The intuition that rare events call for small $p$ is wrong; the sweep selects $p=20\%$.
\item A rank statistic alone does not promote an operational probability, as Section~\ref{sec:calibration} sets out.
\end{itemize}

\subsection{Discrimination is not calibration}
\label{sec:calibration}
An operational product needs both discrimination and calibration. The 2680 MHz channel has the highest stratified \ac{AUC} of the 25, at 0.779. Its Brier skill score is $-0.470$ scored out of sample ($-0.196$ in-sample), so its probabilities are worse than quoting the base rate. Recalibrating in-sample lifts that only to $+0.010$. The channel is therefore not merely miscalibrated. It carries almost no usable probabilistic information, and it is published without fitted weights.

The same signature appears at 872 and 2125 MHz. At 872 MHz the Brier skill is $-0.038$ once in-sample optimism is removed. Both have since been withdrawn from the forecast product. Figure~\ref{fig:reliability} shows discrimination vs. calibration separately. Four channels fall in the region where rank skill is positive and probabilistic skill is not; the fourth, 1865 MHz, never met the eligibility gate. None of them now carries weights. Their discrimination results in Table~\ref{tab:skill} stand on their own.

Scoring such a system needs care. The operational weights are refit on all labelled hours. That is the right choice for a deployed product, but the wrong one for evaluating it. A matched-size control puts the resulting optimism at a median of 0.021 Brier skill. We therefore score the published probabilities against a second weight set fitted on the training period alone. The Brier scores in Table~\ref{tab:skill} and in Figure~\ref{fig:reliability} are that out-of-sample version; scored in-sample they would read a median 0.021 higher, and two channels would cross from negative to positive.

In probability terms the channel set is divided. Against a training-period climatology, the physical prediction adds 0.12 to 0.16 Brier skill on 631, 763, 778, 882, 947, 1842 and 2650 MHz. {It adds only 0.004 to 0.020 on 88, 88.3, 89.3, 162, 162.02, 202 and 460 MHz; that group tracks near-climatology. Both increments are over a seasonal--diurnal climatology, not over the base rate used for the \ac{BSS} column of Table~\ref{tab:skill}.

\begin{figure*}
\includegraphics[width=\textwidth]{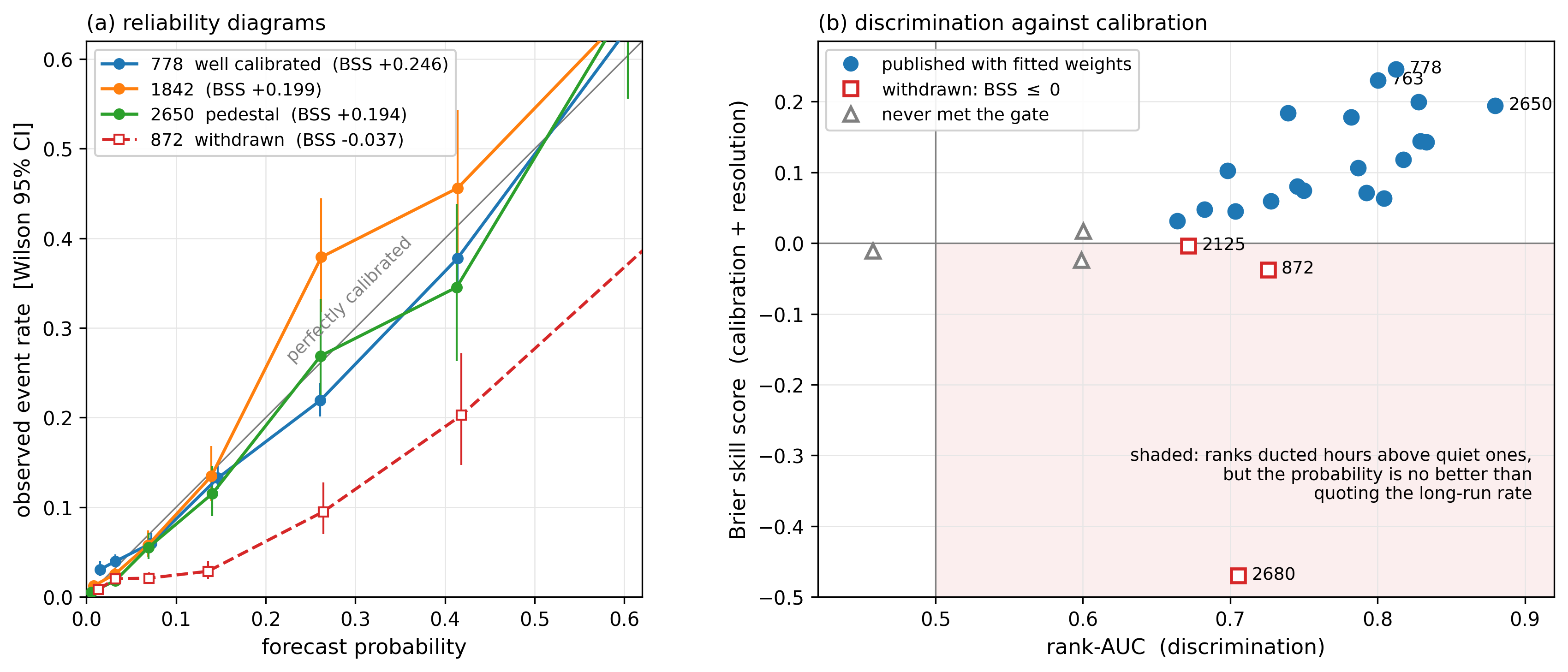}
\caption{Discrimination and calibration are divided, both are scored out of sample against weights fitted on the training period alone. (a) Reliability diagrams with Wilson intervals for a well-calibrated channel, two intermediate cases and one over-confident channel. (b) Every channel on the plane of rank skill against Brier skill; the shaded region, at positive rank skill but negative Brier skill, holds channels that discriminate ducted hours from quiet ones while their probabilities are no better than quoting the base rate. The three channels there carry no fitted weights in the forecast product.}
\label{fig:reliability}
\end{figure*}

\subsection{Terrain data quality}
\label{sec:terrain}
Two terrain defects surfaced during this work, in different datasets. We overlaid \ac{SRTM} terrain on each model's own orography as a routine consistency check on the range-height sections. On the 763 MHz path at azimuth 171$^{\circ}$ the BARRA-C2 profile showed a broad $\sim$650 m terrain rise at about 300 km range. \ac{SRTM} showed nothing there.

Three model-internal diagnostics established that BARRA-C2 does not itself run on the static orographic surface provided by BoM as a companion file with the model data. Its surface pressure at the anomalous cells implies a $\sim$340 m surface instead. The hypsometric height from its own pressure, temperature and geopotential fields matches \ac{SRTM} to a domain-median 13 m. The 950 hPa geopotential is not below-ground-masked, which a 645 m surface would demand. The defect therefore sits in the published static orography ancillary of the AUST-04 collection alone. Within our domain about 1.6\% of land cells disagree with the model's own dynamics by more than 150 m, and by up to $\sim$400 m, in smooth patches over isolated hill clusters. We ruled out envelope orography, grid mismatches and misregistration. We reported the defect to the data custodians. In the pipeline we replaced the static surface with the time-median hypsometric surface wherever the two disagree by more than 100 m. Reprocessing corrected the profile anchor elevation of the affected columns and left every skill metric unchanged to $\pm$0.001, because oceanic corridors dominate the path-median coupling.

The \ac{SRTM} reference might itself have been at fault, so we cross-checked it against the independent Copernicus GLO-90 \ac{DEM} \citep{copernicus_dem}. The two agree to 4--8 m along the affected transect. The BARRA-C2 static file sits 200--300 m above both.

That comparison exposed a second defect, unrelated and in our own propagation inputs. SRTM3 void pixels over salt lakes survive profile extraction and leave spurious pits in 38\% of the transmitter-to-site terrain profiles. Most channels are insensitive to them, because the same corrupted profile enters both the per-timestep loss and its standard-atmosphere reference. An independent audit then traced real damage to one path. A single 577 km profile held a spurious 31 km-deep pit. Its predicted excess of 63--186 dB is physically absurd, and it won the max-over-candidates predictor on every hour for the channels it serves. Its absolute loss was inflated too, which kept it out of the received-power ranking, so only the excess metric suffered. Rebuilding the terrain database from GLO-90, which is void-free over water, raised the excess-metric \ac{AUC} on precisely those channels: 778 MHz from 0.690 to 0.765, 882 MHz from 0.747 to 0.819, 955 MHz from 0.615 to 0.682. All other channels were unchanged, which supports the attribution. All production loss calculations use GLO-90 terrain.

Three lessons came from this experience. Static ancillary fields deserve the same adversarial scrutiny as dynamic data, however reputable the source. Cross-dataset consistency checks are cheap: model hypsometry against static orography, or two independent \acp{DEM} against each other. They locate defects that single-source validation cannot. And a max-over-candidates predictor is fragile to one corrupted input.

\subsection{Limitations}
The \ac{GFS} vertical grid cannot resolve evaporation ducts, or surface ducts shallower than about 100 m. Such ducts trap the lowest paths, so the predictor is blind to a mechanism that does carry signal. Two choices limit the damage. Each profile is anchored at 2 m and at the 0.995-sigma level, which recovers part of the near-surface gradient. Duct layers are floored at 40 m thickness, so unresolved thin features cannot masquerade as ducts; the retracted transmitter-free result of Section~\ref{sec:nulls} shows what happens without that floor. The ERA5 experiment then bounds the residual risk, because 137 vertical levels reproduce the same skill. Vertical truncation is not the binding constraint at this inland site. Over-sea paths may behave differently \citep{zhou2022}, and the near-saturated \ac{AIS} sea segments are consistent with that. It is why the land segment carries the discrimination.

Paths are cropped to the Australia-west region, so transmitters beyond about 1\,000 km are excluded. \ac{AIS} receptions out to 2\,357 km show that propagation at those ranges occurs. Contamination from an excluded source therefore appears as an event the predictor cannot explain. The licence register misses spectrum-licensed sites, meaning we only know there are transmitters, but not where they are, with the same consequence. The 2.6 GHz blocks are the clear case: they are spectrum-licensed and invisible in the register, which is why their common tower had to be established empirically (Section~\ref{sec:imt}). Apparatus licences cover the mobile sites that dominate here, so the gap is small in practice. Both omissions bias the measured skill downward rather than upward.

Driving a statistical propagation model per timestep is a hybrid approach. P.452 and P.1812 return the loss not exceeded for a stated percentage of an average year, so a per-hour value is not a physical prediction of the loss in that hour, and the time percentage acts as a calibration constant. We apply two mitigations to constrain the consequence. We score by rank, which is invariant to any monotone rescaling of the predictor. The operating point also sits on a plateau, flat from $p=20\%$ to $p=50\%$, so the choice is not critical. Absolute probabilities come from the learned layer instead, and its calibration is reported separately in Section~\ref{sec:calibration}.

At \ac{FM} frequencies the achievable contrast is smaller. P.1812 floors the time percentage at 1\%, and the \ac{VHF} duct response is intrinsically weaker: the loss spans about 25 dB across the plausible range of $\Delta N$ at 88 MHz, against about 75 dB on the 956 MHz test path. A weaker \ac{FM} result would therefore be expected even where ducting does occur. We do not rest the \ac{FM} attribution on the propagation model alone for that reason. The mechanism exclusions and the 162 MHz \ac{AIS} positive control carry it.

The event label uses a centred rolling baseline, which is not causal. That is correct for a retrospective validation, and we note this as a deployment caveat. It also does not inflate the result: relabelling causally moves the stratified \ac{AUC} by a median of $+0.004$, and 20 of the 25 channels score higher that way. The deployed system runs a trailing baseline.

P.452's effective-earth-radius factor is singular at $\Delta N = 157$ N-units per km, the super-refraction threshold, so the recommendation cannot be evaluated inside a duct at all. Averaging the gradient over the lowest kilometre and then along the path keeps the model well away from that limit -- over 21\,000 grid columns spanning strong-event and quiet hours, the largest path-column value we encounter is 120 N-units per km -- but it means the ducting itself is never carried by $\Delta N$. It is carried by the anomalous-propagation term that $p$ selects, which is why $p$ has to be calibrated rather than assumed.

\subsection{Learned contamination likelihood}
\label{sec:ml}
A learned model extracts more skill from the same physical inputs. The split is strictly temporal: train on 2022--2024, test on 2025--2026. A pooled \ac{GBM} combines the analytical prediction with the raw duct-field diagnostics, the path $\Delta N$ statistics and diurnal--annual harmonics. It reaches a mean test \ac{AUC} of 0.795 across all 25 channels, against 0.688 for the same model restricted to the harmonics (Table~\ref{tab:ml} and Figure~\ref{fig:mlbaselines}). Adding the corridor feature of Section~\ref{sec:corridor} raises it to 0.806.

Restricted to the 16 channels on which the \emph{analytical} predictor is itself skilled, the same two numbers read 0.785 without the corridor feature and 0.799 with it, at Brier skill 0.103 and 0.133. That the restricted mean is the lower of the two is not a weakness of the subset but a property of the gate, which is set on the analytical \ac{AUC} and not on the learned one. The nine excluded channels average 0.819 under the learned model (above the skilled mean) and include the two highest-scoring channels in the set, 162 and 162.02 MHz at 0.892 and 0.895, where the analytical predictor returns essentially chance. The learned model therefore recovers usable skill on channels where the deterministic physical chain does not, which is a result in its own right rather than an artefact of averaging.

We use ablation as the figure of merit (Figure~\ref{fig:mlablation}). Restricting the pooled model to the climatology harmonics alone drops it from 0.795 to 0.688, while restricting it to the physical features costs only 0.010, to 0.786. That physics-only advantage survives stratification within month and within hour, so the physics carries real intra-seasonal and intra-diurnal discrimination and is not a season proxy. The model exploits non-linear coupling of existing physical features. It does not discover new physics. The pooled per-channel values come from one fit by design, which is the right choice for the sparse channels.

\begin{figure}[t]
\includegraphics[width=\columnwidth]{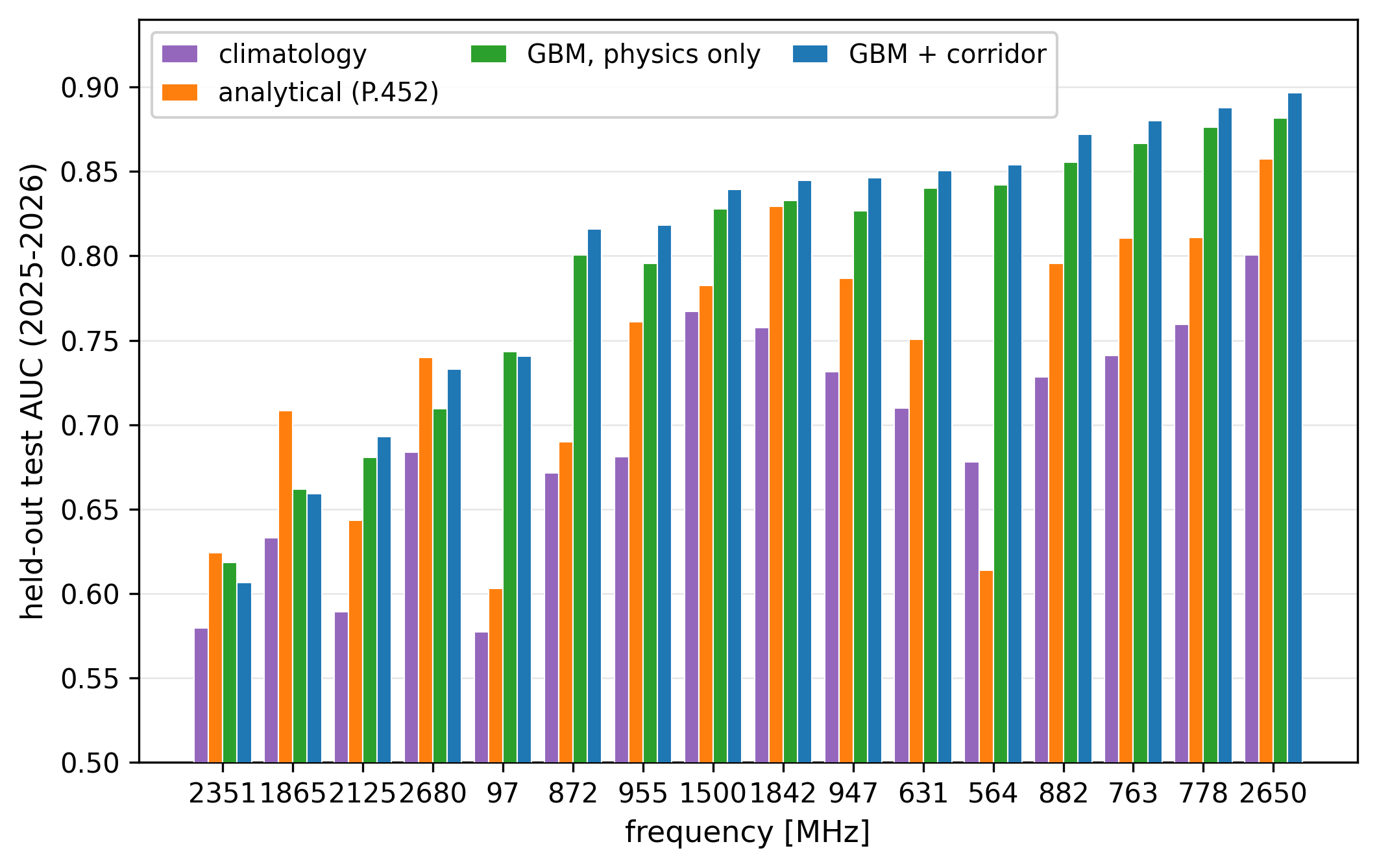}
\caption{Per-channel test \ac{AUC} for the predictor ladder on the 2025--2026 test period, over the analytically-skilled channels: the diurnal--annual climatology, the analytical P.452 prediction, and the pooled \ac{GBM} without and with the corridor feature. Persistence and the per-channel logistic baseline are not shown; they were not recomputed on the production basis, and drawing them from the superseded run would mix bases within one panel.}
\label{fig:mlbaselines}
\end{figure}

\begin{figure}[t]
\includegraphics[width=\columnwidth]{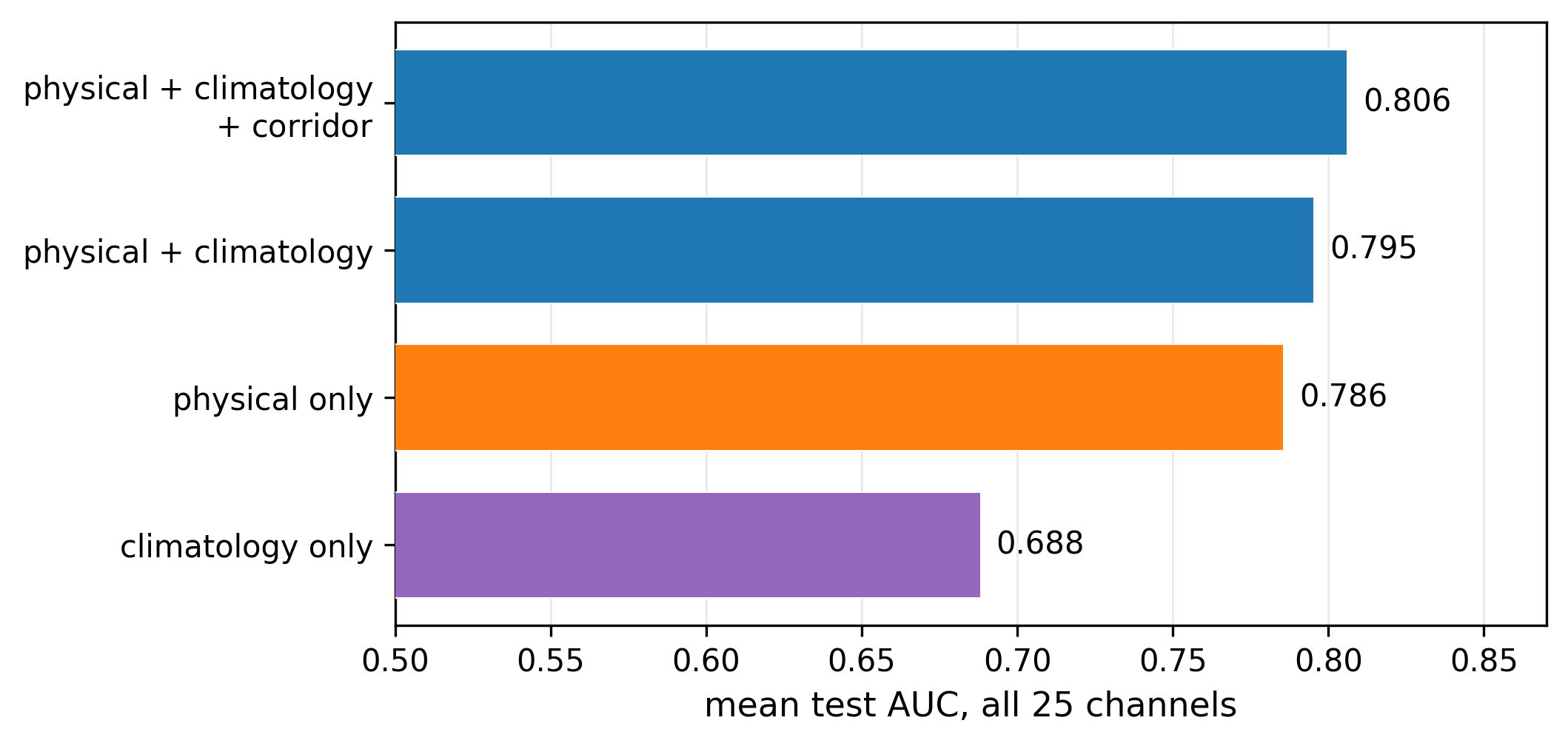}
\caption{Feature ablation of the learned contamination-likelihood model, on the same fit as Table~\ref{tab:ml} and averaged over the same 25 channels. The top two bars are Table~\ref{tab:ml}'s first two rows and the bottom bar its third; the third bar drops the climatology harmonics. Dropping climatology costs 0.010 in mean \ac{AUC}, whereas dropping the physical features costs 0.107: the gain over the analytical predictor is physics-driven. The physical-only set retains the two persistence terms, being the complement of the climatology harmonics rather than a pure-propagation set.}
\label{fig:mlablation}
\end{figure}

\begin{table}[t]
\caption{Skill of the learned contamination-likelihood model on the 2025--2026 test period, averaged over all 25 channels; all three rows are the same pooled gradient-boosted ensemble, differing only in which input feature groups it is given. The model is trained on the v16 era only, so the v15 era remains available as an out-of-era check.}
\label{tab:ml}
{\tablefont\begin{tabular}{@{\extracolsep{\fill}}lc}
\toprule
Model & \ac{AUC}\\
\hline
Learned model, physical features $+$ climatology $+$ corridor & 0.806\\
Learned model, physical features $+$ climatology & 0.795\\
Learned model, climatology only & 0.688\\
\botrule
\end{tabular}}
\end{table}

\subsection{Duct-corridor connectivity}
\label{sec:corridor}
The resolution ladder of Section~\ref{sec:validation} suggested that the path-median $\Delta N$ coupling discards the corridor structure finer models resolve. A corridor-connectivity metric recovers part of it. For every transmitter-to-site path and hour we extract the duct layers of each column along the path, floored at 40 m thickness. We then score the path three ways: by its ducted fraction, by the longest height-overlapping chain of layers, and by whether such a chain reaches the observatory at all. The third is a yes/no test, and we call a path connected when it passes. A graded variant replaces that yes/no answer with a strength, weighting the coupling along the chain by distance decay. The aggregation runs over the same transmitter population as Section~\ref{sec:attribution}: 9\,479 licensed transmitter-channel entries, de-duplicated to 9\,404 unique channel-sites at $\sim$100 m coordinate precision, with co-located licences merged and the maximum \ac{EIRP} retained.

The connectedness test is itself a negative result. Standalone it discriminates events at only 0.53--0.59, measured on the twelve channels of the corridor development set. That is below the regional land duct fraction of 0.628 and far below the analytical predictor, and it carries no measurable weight in the operational logistic model. Two refinements change that.

The first corrects a bias in the duct decision. \ac{GFS} over-detects ducts at sea and under-detects them over land, so the error is opposite-signed by surface type. A surface-conditioned threshold fixes it, rate-matched to ERA5 land duct occupancy and using no event labels. The operational value is $\tau_{\mathrm{land}} = 49.07$ M-units km$^{-1}$. It is a calibrated ``ERA5-would-see-a-duct'' decision, not a physical duct criterion. Figure~\ref{fig:corridorprofiles} shows it working along real corridors, recovering land duct layers that the physical onset contour misses.

The second refinement replaces the yes/no test with the graded coupling. On the 2025--2026 test period that coupling reaches 0.737 standalone. As an added feature its median per-channel gain to the boosted model is $+0.009$, individually significant on 14 of the 25 channels (Figure~\ref{fig:corridorskill}). The deployed operational model gains $+0.018$, from 0.737 to 0.756 over its skilled channels.

\begin{figure}[t]
\includegraphics[width=\columnwidth]{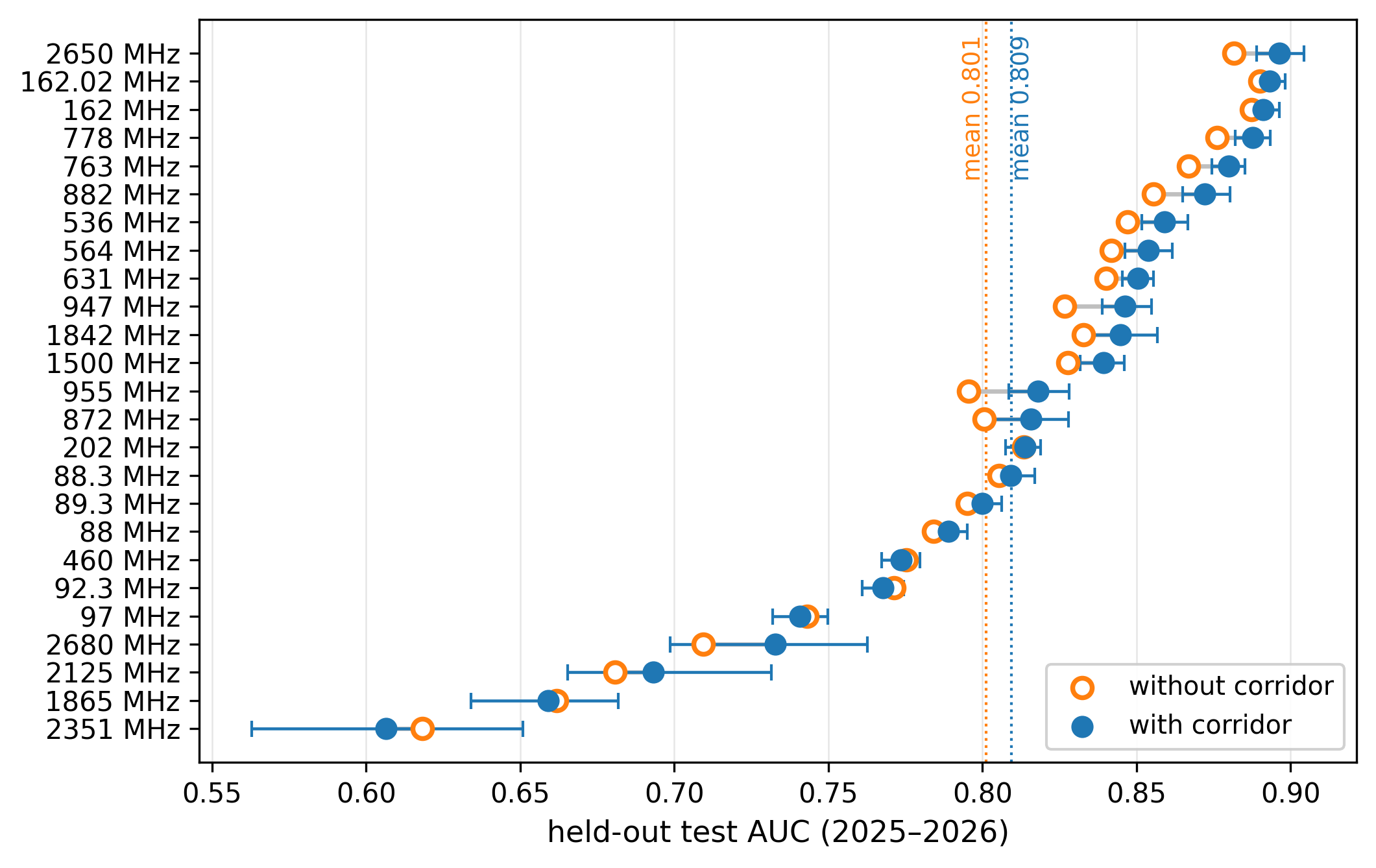}
\caption{Per-channel effect of the corridor-connectivity feature on the pooled \ac{GBM}: test \ac{AUC} without (open) and with (filled) the corridor coupling, with paired block-bootstrap confidence intervals.}
\label{fig:corridorskill}
\end{figure}

\begin{figure*}
\includegraphics[width=\textwidth]{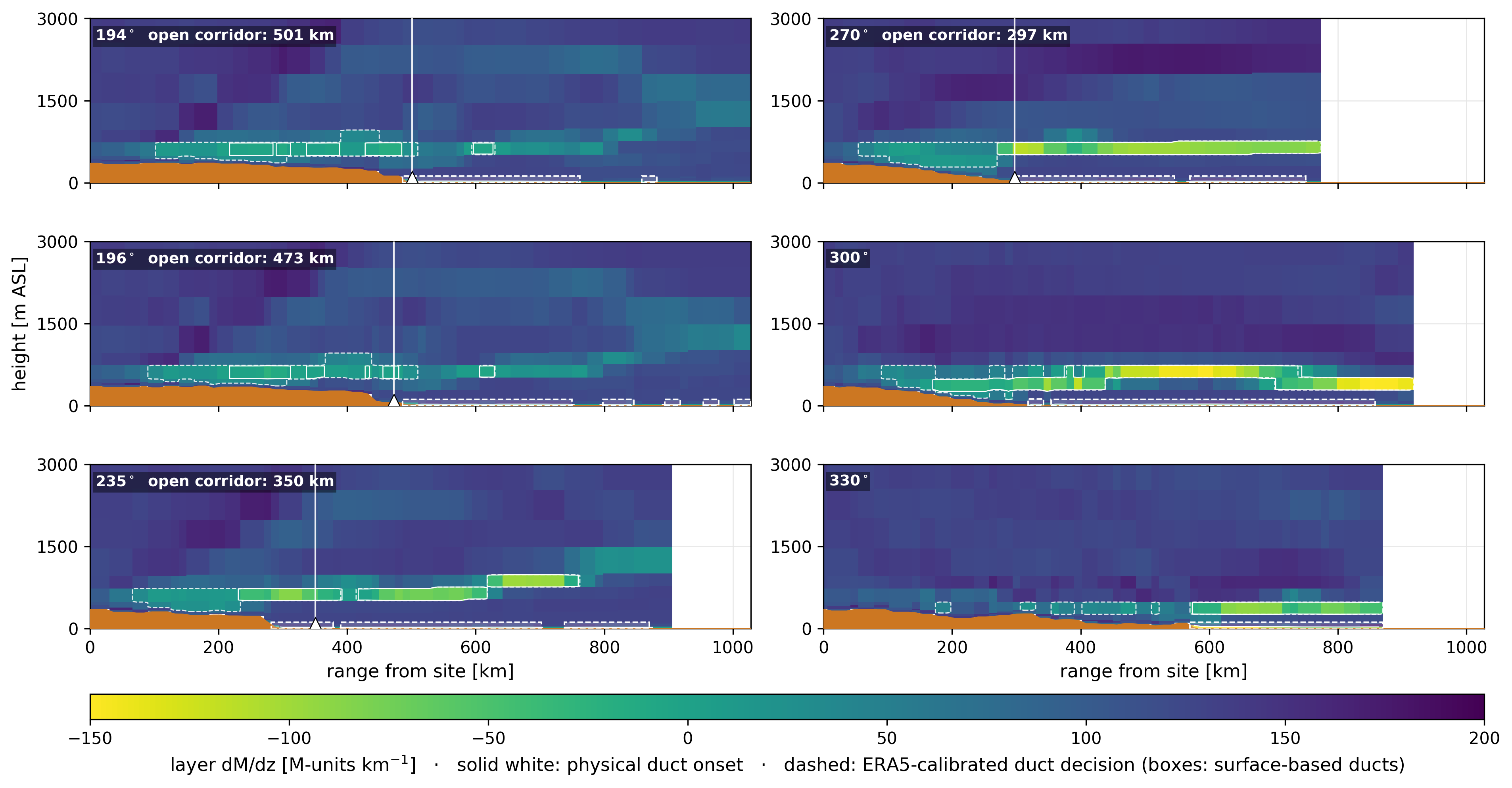}
\caption{Range-height sections of the layer gradient $\mathrm{d}M/\mathrm{d}z$ along six bearings from the observatory at the 2023 February 5 21:00 UT case-study hour. Solid contours mark the physical duct onset ($\mathrm{d}M/\mathrm{d}z < 0$); dashed contours mark the ERA5-calibrated duct decision ($\tau_{\mathrm{land}}$), which recovers the land duct structure that \ac{GFS} under-detects; dashed boxes along the ground mark surface-based ducts. Four bearings (194$^{\circ}$, 196$^{\circ}$, 235$^{\circ}$, 270$^{\circ}$) carry open corridors at this hour, with the corridor extent and the furthest coupled transmitter marked; the remaining two are duct-active but unlocked bearings. Terrain is ochre colour filled.}
\label{fig:corridorprofiles}
\end{figure*}

The calibration rests on a measurable difference between the two models' duct fields, and that difference is a land phenomenon. Over 2\,880 common hours on a common 0.25$^{\circ}$ grid, ERA5 diagnoses a ducting layer over land in 10.4\% of cell-hours. \ac{GFS} diagnoses one in 4.5\%, a factor of 2.3. The deficit grows to roughly fourfold, with a range of 2.4 to 8.5, in the rate at which a duct is present within about 50 km of the observatory; that ratio is computed on the pre-correction site position, the raw-\ac{GFS} arm never having been recomputed on the surveyed one. At sea the bias reverses: ERA5 reads 56.0\% against 73.1\% for \ac{GFS}. \ac{GFS} under-representation is therefore a statement about land alone.

After the rate-match the two agree on the near-site detection rate, 11.4\% against 10.2\%. That agreement checks the calibration; it does not validate \ac{GFS} against ERA5. The threshold was fitted by matching land occupancy on these same hours, the detection rate is itself an occupancy statistic, and three of the four months are the fitting sample. The pooled figure also hides opposite-signed seasonal errors that cancel, because the corrected \ac{GFS} rate runs high in summer and low in winter. Geometry is not neutral either, and the two models respond to it with opposite sign, so the comparison cannot be assumed robust to the site position. The independent evidence for the corridor work is its test-period skill, not this comparison.

The threshold survives a seasonal-generality check. Rate-matched refits over paired GFS--ERA5 months run from about $+20$ M-units km$^{-1}$ in winter to about $+58$ in summer, tracking ERA5 land duct occupancy of 3--16\%. Summer refits are stable across years. The winter fits are noisy, because occupancy is low. Corridor skill is flat from $+35$ to $+58$, and the operational value sits on that plateau. Duct events concentrate in the high-occupancy season, so we retain the fixed calibration. Only the winter extreme would halve the corridor gain.

\subsection{Operational implications}
The pipeline runs from openly archived \ac{GFS} forecast data, and skill is flat through three days of lead (Section~\ref{sec:validation}). The same diagnostics computed from forecast hours therefore give useful interference forecasts up to three days ahead. Duct-aware scheduling can move sensitive observations to unaffected frequencies or times. Flagged periods can also inform post-hoc data quality assessment.

A real-time forecast built on this method is already in operation, refreshed on every 6-hourly \ac{GFS} cycle, available on the web\footnote{\url{https://www.atnf.csiro.au/observers/DFS/}} and published as a machine-readable time series for ingestion into the observatory's operational database, for consumption by scheduling software. The live service runs the surveyed position and all 25 channels, 19 of them with fitted weights, using a trailing rather than a centred baseline as a causal service requires, and runs the corridor-augmented weights of Section~\ref{sec:corridor}. Its map overlays the duct-coupled transmitter sites: coupling rings appear in $\sim$1.2\% of hours, and those hours carry contamination odds ratios of 2.6--8.5.

Hindcasts run archival \ac{GFS} cycles through the full operational stack and test it against independent ground truth (Figure~\ref{fig:hindcast}, also available online\footnote{\url{https://www.atnf.csiro.au/observers/DFS/hindcast/2022-02-24_18z/}\\\url{https://www.atnf.csiro.au/observers/DFS/hindcast/2022-11-28_18z/}\\\url{https://www.atnf.csiro.au/observers/DFS/hindcast/2023-02-03_12z/}\\\url{https://www.atnf.csiro.au/observers/DFS/hindcast/2023-02-12_12z/}\\}). Take the 2022 November 28 18Z cycle. It successfully predicted the November 30 event, during which the site \ac{LTE} scanner decoded 100 cells from 13 sites in the wheatbelt east of Perth. That cycle placed the event at roughly two days of lead. The 763 MHz contamination probability averaged 0.93 inside the decode window and peaked at 0.98, against 0.35 outside it. The coupled-site rings extend inland across the wheatbelt, matching the decoded geography. Further hindcasts called both peaks of a multi-day Exmouth episode from a single cycle, at forecast hours 22--29 and 69--79, the latter close to three days of lead, and the 2023 February 5 and 14 events at probabilities up to 0.91. The decodes are independent of the spectrum monitor that trained the system, so these are end-to-end verifications.

The corridor fields for that frame qualify the demonstration rather than confirm it (Figure~\ref{fig:corridorwheat}). At the forecast peak, 9 of the 13 decoded sites lay on paths carrying a height-continuous ducted chain to the observatory. Across the seven-hour decode window 12 of 13 did so at some hour. The exception is noteworthy. It is Burracoppin, the second-largest decode site (18 decodes, 161.9$^{\circ}$, 554 km), whose coupling peaks at 0.236 against the 0.25 detection threshold and so never opens. Merredin, the largest (28 decodes) and only 1.7$^{\circ}$ away, crosses only marginally and only briefly, peaking at 0.272. Between them these two sites carry 46 of the 100 decodes on a corridor the model rates as barely passable. The section on that bearing shows a broken duct, a surface-based layer to about 250 km and then a gap. The bearing carrying the second cluster shows a continuous elevated layer instead. Either the duct is too shallow or too narrow for a 0.25$^{\circ}$ grid, or the coupling threshold is too strict for that geometry. The metric therefore locates most of the decoded geography but not all of it, and at the plotted hour it accounts for fewer than half the decodes by count.

\begin{figure*}
\includegraphics[width=\textwidth]{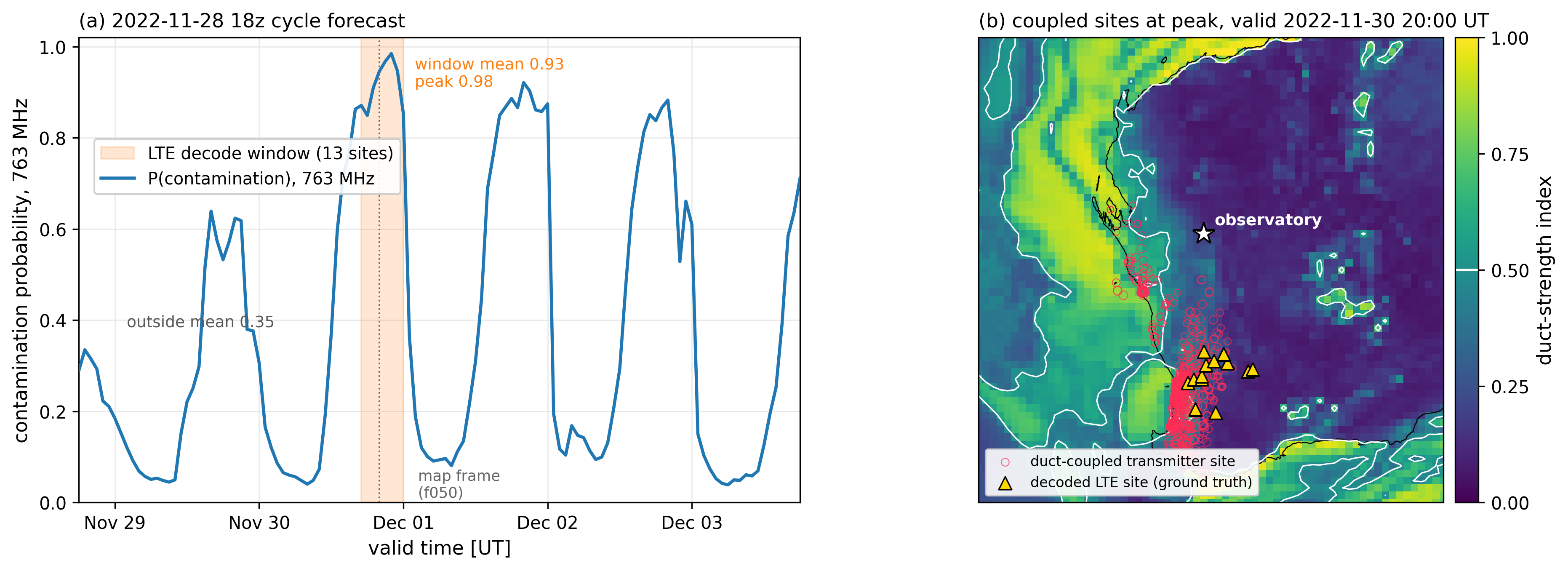}
\caption{Hindcast of the 2022 November 30 wheatbelt interference event from the 2022 November 28 18Z \ac{GFS} cycle through the full operational forecast stack. (a) Contamination probability at 763 MHz across the forecast horizon; the shaded band marks the window in which the site \ac{LTE} scanner decoded 100 cells from 13 wheatbelt sites east of Perth. (b) The duct-strength index and duct-coupled transmitter sites (circles) at the forecast peak, with the decoded \ac{LTE} sites (triangles) as independent ground truth: the coupling extends inland across the wheatbelt, matching most of the decoded geography, with the qualification quantified in Section~\ref{sec:corridor}.}
\label{fig:hindcast}
\end{figure*}
\begin{figure*}
\includegraphics[width=\textwidth]{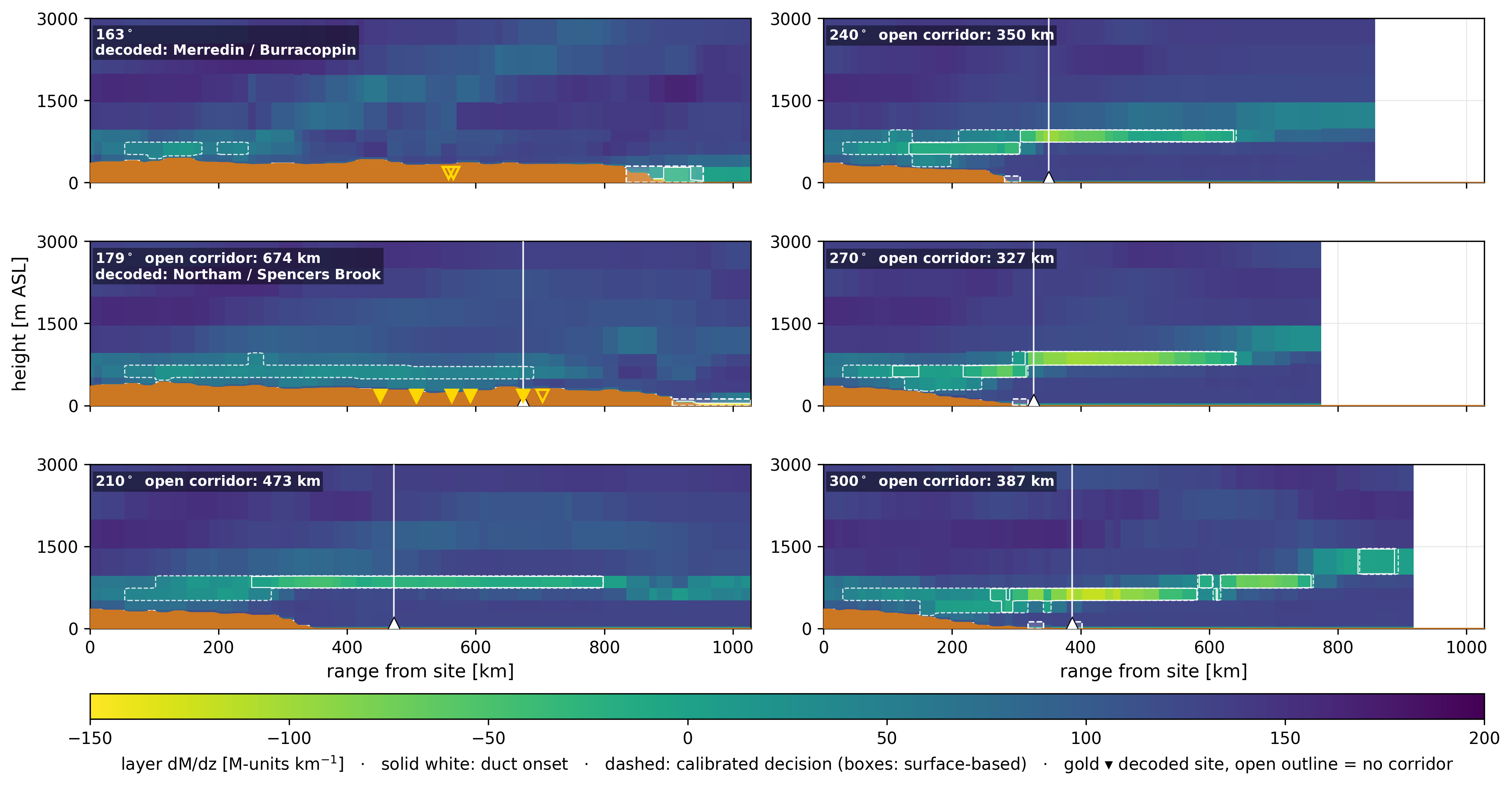}
\caption{Range-height sections along six bearings from the observatory at the 2022 November 30 20:00 UT forecast peak, the frame of Figure~\ref{fig:hindcast}(b), under the operational calibration. Conventions follow Figure~\ref{fig:corridorprofiles}. Gold triangles mark decoded \ac{LTE} sites; an open outline means the site sits on a path with no open corridor at this hour. The 163$^{\circ}$ bearing carries the two highest-decode sites on a broken duct, while the 179$^{\circ}$ bearing carries the second cluster on a continuous elevated layer.}
\label{fig:corridorwheat}
\end{figure*}

\section{Conclusions}
\label{sec:conclusions}

Tropospheric ducting delivers detectable terrestrial interference into the \ac{ARQZWA} on thousands of hours per channel, with a strong summer maximum. We compute duct diagnostics from archival \ac{GFS} 0.25$^{\circ}$ data and couple them to per-transmitter basic transmission loss, using ITU-R P.452-16 and P.1812-6 over Copernicus GLO-90 terrain. Across 25 monitored channels the median stratified \ac{AUC} is 0.724, and 24 of the 25 intervals exclude chance. Scoring within month and three-hour strata is essential to that claim. A pooled statistic would credit 12 of the channels for a seasonal and diurnal cycle they share with the predictor. A time-percentage calibration fixed on four pilot months transfers to the full record without retuning. Skill holds through three days of forecast lead. A validation era from a different \ac{GFS} model version, never trained on, reproduces the per-channel result.

Independent \ac{LTE} cell decoding confirms the events and the attributed sources out to 589 km. Self-located \ac{AIS} ship receptions confirm that the predicted ducts gate long-range \ac{VHF} reception, though decodes lag the onset of telescope-affecting propagation by a median of four hours. Two channels at 2.6 GHz were recovered from self-generated telescope \ac{RFI} by discriminating on bandwidth rather than amplitude. Any observatory monitoring a band it also pollutes can adopt that method, though ideally such pollution should be prevented from the onset. The land mobile channel at 460 MHz is a well-powered null, the only one of the 25 whose interval contains chance. The \ac{FM} channels at 88, 88.3 and 89.3 MHz retain weak but non-zero skill, far below the duct-driven channels, which we read as near-threshold reception of local low-power sources modulated by ordinary refractive variability: a result about attribution rather than physics.

The result is a practical, site-agnostic recipe. Open \ac{NWP} data, an open licence register, open terrain data and open-source propagation implementations suffice to predict ducting-borne interference at a radio-quiet site. A pooled \ac{GBM} on the same physical inputs improves on the analytical prediction. A duct-corridor connectivity feature adds a further increment, individually significant on 14 of 25 channels. Those weights drive the observatory's live forecast service, and archival hindcasts verify it end-to-end against independently decoded events at leads of one to three days. Discrimination and calibration remain separate claims. Several channels discriminate well but carry no usable probabilistic information, and their operational products are unvalidated until refit and scored out of sample.

\clearpage
\section*{Acknowledgements}
We acknowledge the Wajarri Yamaji People as the Traditional Owners and native title holders of Inyarrimanha Ilgari Bundara, the CSIRO Murchison Radio-astronomy Observatory site. \ac{GFS} data were obtained through the NOAA Open Data Dissemination program on AWS. Transmitter licensing data are from the \ac{ACMA} Register of Radiocommunications Licences. Terrain data are from the ESA Copernicus GLO-90 digital elevation model and the NASA Shuttle Radar Topography Mission. Ionosonde data are from the Bureau of Meteorology Space Weather Services.

\section*{Data Availability}
The code and derived data used for this study is not publicly available. The \ac{GFS} data is public and available here \url{https://registry.opendata.aws/noaa-gfs-bdp-pds/}. The \ac{RRL} data is also public and available here \url{https://www.acma.gov.au/register-radiocommunications-licences-rrl}. CSIRO's live Ducting Forecasting System is available here: \url{https://www.atnf.csiro.au/observers/DFS/}

\bibliographystyle{paslike}
\bibliography{ducting}

\end{document}